\documentclass[twocolumn]{aastex63}

\newcommand{\degree}{\ensuremath{^\circ}}
\newcommand{\degrees}{\ensuremath{^\circ}}
\newcommand{\amin}{\ensuremath{^\prime}}
\newcommand{\msun}{\ensuremath{M_{\odot}}}
\newcommand{\fermi}{\textit{Fermi}}

\newcommand{\rsun}{\ifmmode R_{\odot}\else$R_{\odot}$\fi}
\newcommand{\asec}{\ifmmode^{\prime\prime}\else$^{\prime\prime}$\fi}

\newcommand{\xmm}{{\it XMM-Newton}}
\newcommand{\gray}{$\gamma$-ray}

\newcommand{\dm}{\,pc cm$^{-3}$}

\newcommand{\edot}{$\dot{\rm E}$}

\newcommand{\mspa}{J1810+1744} %BW
\newcommand{\mspb}{J0023+0923} %BW
\newcommand{\mspc}{J2215+5135} %RB
\newcommand{\mspd}{J1124$-$3653} %BW
\newcommand{\mspe}{J2129$-$0429} %RB
\newcommand{\bwa}{B1957$+$20} %Fruchter et al. 1988
\newcommand{\bwc}{J2051$-$0827} %Stappers et al. 1996
\newcommand{\spia}{J1023$+$0038} %Archibald et al. 2009
\newcommand{\maura}[1]{}

\shorttitle{GBT 350 MHz Survey of Unassociated Fermi LAT Sources}
\shortauthors{Bangale et al.}

\begin{document}
\title{A 350-MHz Green Bank Telescope Survey of Unassociated \\ \fermi\ LAT Sources:  Discovery and Timing of Ten  Millisecond Pulsars}
%---------------------------------------------------------------------------------

\author[0000-0002-3886-3739]{P. Bangale}
\affiliation{Department of Physics and Astronomy, West Virginia University, Morgantown, WV 26506, USA}
\affiliation{Department of Physics and Astronomy, Bartol Research Institute, University of Delaware, Newark, DE 19716, USA}

\thanks{Corresponding authors: \\ P. Bangale (\url{priyadarshini.bangale@gmail.com}), \\ M. Kerr (\url{matthew.kerr@nrl.navy.mil)}}

\author[0000-0002-6287-6900]{B. Bhattacharyya}
\affiliation{National Centre for Radio Astrophysics, Tata Institute of Fundamental Research, Pune 411 007, India}
 
\author[0000-0002-1873-3718]{F. Camilo}
\affiliation{South African Radio Astronomy Observatory (SARAO), 2 Fir Street, Observatory, Cape Town, 7925, South Africa}
 
\author[0000-0003-4355-3572]{C. J. Clark}
\affiliation{Max Planck Institute for Gravitational Physics (Albert Einstein Institute), D-30167 Hannover, Germany}
\affiliation{Leibniz Universit\"at Hannover, D-30167 Hannover, Germany}
 
\author[0000-0002-1775-9692]{I. Cognard}
\affiliation{Station de radioastronomie de Nan\c{c}ay, Observatoire de Paris, CNRS/INSU, F-18330}
 
\author[0000-0002-2185-1790]{M. E. DeCesar}
\affiliation{College of Science, George Mason University, Fairfax, VA 22030, resident at Naval Research Laboratory, Washington, DC 20375, USA}

\author{E.~C. Ferrara}
\affiliation{Department of Astronomy, University of Maryland, College Park, MD 20742}
\affiliation{Center for Research and Exploration in Space Science and Technology, NASA/GSFC, Greenbelt, MD 20771}
\affiliation{NASA Goddard Space Flight Center, Greenbelt, MD 20771, USA}

\author{P. Gentile}
\affiliation{Department of Physics and Astronomy, West Virginia University, Morgantown, WV 26506, USA}

\author[0000-0002-9049-8716]{L.~Guillemot}
\affiliation{Laboratoire de Physique et Chimie de l'Environnement et de l'Espace -- Universit\'e d'Orl\'eans / CNRS, F-45071 Orl\'eans Cedex 02, France}
\affiliation{Station de radioastronomie de Nan\c{c}ay, Observatoire de Paris, CNRS/INSU, F-18330 Nan\c{c}ay, France}
 
\author[0000-0003-2317-1446]{J.~W.~T. Hessels}
\affiliation{ASTRON, the Netherlands Institute for Radio Astronomy, Postbus 2, 7990 AA Dwingeloo, The Netherlands}
\affiliation{Anton Pannekoek Institute for Astronomy, University of Amsterdam, Science Park 904, 1098 XH Amsterdam, The Netherlands}

\author{T.~J. Johnson}
\affiliation{College of Science, George Mason University, Fairfax, VA 22030, resident at Naval Research Laboratory, Washington, DC 20375, USA}

\author[0000-0002-0893-4073]{M. Kerr}
\affiliation{Space Science Division, U.S. Naval Research Laboratory, Washington, DC 20375, USA}

\author[0000-0001-7697-7422]{M. A. McLaughlin}
\affiliation{Department of Physics and Astronomy, West Virginia University, Morgantown, WV 26506, USA}
\affiliation{Center for Gravitational Waves and Cosmology, Chestnut Ridge Research Building, Morgantown, WV 26505, USA}

\author[0000-0002-5775-8977]{L. Nieder}
\affiliation{Max Planck Institute for Gravitational Physics (Albert Einstein Institute), D-30167 Hannover, Germany}
\affiliation{Leibniz Universit\"at Hannover, D-30167 Hannover, Germany}

\author[0000-0001-5799-9714]{S.~M. Ransom}
\affiliation{National Radio Astronomy Observatory, 520 Edgemont Road, Charlottesville, VA 22903-2475, USA}
 
\author[0000-0002-5297-5278]{P.~S. Ray}
\affiliation{Space Science Division, U.S. Naval Research Laboratory, Washington, DC 20375, USA}

\author{M.~S.~E. Roberts}
\affiliation{Eureka Scientific Inc, Oakland, CA. USA}

\author{J. Roy}
\affiliation{National Centre for Radio Astrophysics, Tata Institute of Fundamental Research, Pune 411 007, India}

\author{S. Sanpa-Arsa}
\affiliation{National Astronomical Research Institute of Thailand (Public Organization), 260 Moo 4, T. Donkaew, A. Maerim, Chiang Mai, 50180, Thailand}

\author[0000-0002-3649-276X]{G.~Theureau}
\affiliation{Laboratoire de Physique et Chimie de l'Environnement et de l'Espace -- Universit\'e d'Orl\'eans / CNRS, F-45071 Orl\'eans Cedex 02, France}
\affiliation{Station de radioastronomie de Nan\c{c}ay, Observatoire de Paris, CNRS/INSU, F-18330 Nan\c{c}ay, France}
\affiliation{LUTH, Observatoire de Paris, Universit\'e PSL, Universit\'e Paris Cit\'e, CNRS, 92195 Meudon, France}

\author[0000-0002-4013-5650]{M.~T. Wolff}
\affiliation{Space Science Division, U.S. Naval Research Laboratory, Washington, DC 20375, USA}

%---------------------------------------------------------------------------------
\begin{abstract}
We have searched for radio pulsations towards 49 \fermi\ Large Area Telescope (LAT) 1FGL Catalog $\gamma$-ray sources using the Green Bank Telescope at 350 MHz. We detected 18 millisecond pulsars (MSPs) in blind searches of the data; 10 of these were discoveries unique to our survey.
Sixteen are binaries, with eight having short orbital periods $P_B < 1$\,d. No radio pulsations from young pulsars were detected, although three targets are coincident with apparently radio-quiet $\gamma$-ray pulsars discovered in LAT data. Here, we give an overview of the survey and present radio and $\gamma$-ray timing results for the 10 MSPs discovered. These include the only isolated MSP discovered in our survey and six short-$P_B$ binary MSPs. Of these, three have very low-mass companions ($M_c \ll 0.1\msun$) and hence belong to the class of black widow pulsars. Two have more massive, non-degenerate companions with extensive radio eclipses and orbitally modulated X-ray emission consistent with the redback class. Significant $\gamma$-ray pulsations have been detected from nine of the discoveries. This survey and similar efforts suggest that the majority of Galactic $\gamma$-ray sources at high Galactic latitudes are either MSPs or relatively nearby non-recycled pulsars, with the latter having on average a much smaller radio/$\gamma$-ray beaming ratio as compared to MSPs. It also confirms that past surveys suffered from an observational bias against finding short-$P_B$ MSP systems.

\end{abstract}

%---------------------------------------------------------------------------------
\section{Introduction}

One of the major results to come out of over a decade of observations with the \textit{Fermi} Gamma-ray Space Telescope (hereafter \fermi) is that the majority of  Galactic high-energy ($E\sim1$\,GeV) $\gamma$-ray sources so far detected are pulsars. At low Galactic latitudes ($b \la 10^{\circ}$), it had been suspected since the EGRET era that most discrete sources might be young ($\tau \la 10^{5.5}$~yr), spin-powered pulsars \citep{rob05}.  Before the launch of \fermi, the most likely population of Galactic $\gamma$-ray sources at higher Galactic latitudes was thought to be nearby ($d \la 2$~kpc) middle-aged pulsars like PSR B1055$-$52 and Geminga, perhaps associated with the Gould Belt of nearby star-forming regions \citep{ggh+05}. Millisecond pulsars (MSPs, which, for the purpose of this paper, are defined as having spin periods $P < 10$~ms) were considered a possibly important class as well, despite only one marginal detection with EGRET \citep{khv+00}. Radio pulsar surveys targeting EGRET source error boxes \citep{ns97,cml05,crh+06} were notoriously unsuccessful, with only three pulsars discovered in these surveys---the MSPs J0751$+$1807 \citep{Lundgren95} and J1614$-$2230 \citep{hrr+05,dpr+10}, and the young PSR J1028$-$5819 \citep{kjk+08}---that were plausibly the counterparts of the $\gamma$-ray source targeted \citep{aaa+10a}. Very deep radio searches of low-latitude extended X-ray sources discovered within EGRET error boxes were somewhat more efficient at discovering potential $\gamma$-ray pulsars. Two young and energetic pulsars, PSR J2021+3651 \citep{rhr+02} and PSR J2229+6114 \citep{hcg+01} were discovered this way, both of which proved to be powering their respective coincident $\gamma$-ray sources \citep{hcg+08,ppp+09}.

With the launch of the \fermi\ satellite, it was quickly determined that MSPs were efficient $\gamma$-ray emitters, with eight MSP detections reported within the first few months of operations \citep{aaa+09b}. In addition, blind searches of the \fermi\ Large Area Telescope \citep[LAT,][]{Atwood09} data soon discovered 16 young pulsars, three of which were at mid-Galactic latitudes \citep[$10^\circ < |b| < 30^\circ$][]{aaa+09a}. The 46 $\gamma$-ray pulsars reported in the First \fermi-LAT Pulsar Catalog \citep{aaa+10a} include 15 pulsars at $|b| > 5^{\circ}$. Of these, eight are MSPs, four are young radio pulsars, and three are detected only in $\gamma$ rays. It was therefore expected that deep radio searches of unassociated LAT sources at mid latitudes should result in a mix of MSPs and young pulsars. 

A sensitive 820-MHz radio search with the 100-m Green Bank Telescope (GBT) of 25 of the brightest LAT sources not associated with previously known sources of potential $\gamma$ rays (i.e., energetic pulsars or blazars) discovered no young pulsars. However, out of the eight sources searched at  $|b| > 5^{\circ}$, three were found to be binary radio MSPs \citep{rrc+11}. One of these four, PSR J2214+3000, is in a tight ($P_B \sim 10$~hr) orbit around a very low-mass (minimum companion mass $M_c \sim 0.02 M_{\odot}$) companion and exhibits regular eclipses, becoming at that time only the fourth known ``black widow" system discovered in the Galactic field (outside of a globular cluster).

Between the discovery of the first MSP in 1982, the isolated 1.6-ms PSR B1937+21 \citep{bkh+82}, and the launch of \fermi\ in 2008, psrcat\footnote{http://www.atnf.csiro.au/research/pulsar/psrcat/} \citep{psrcat} lists the discovery of approximately 
70 MSPs in the Galactic field. In comparison, since data from \fermi\ became publicly available in 2009, over 100 MSPs (including the ones reported here) were discovered in the Galactic field by targeting \fermi\ LAT sources \citep{cgj+11,kjr+11,gfc+12,kcj+12,rap+12,bgc+13,ckr+15,cck+16,pleun17,den21, bhat2013, bhat21, wlc+21, cbb+23}. 
Surprisingly, only one young $\gamma$-ray pulsar has been discovered through targeted radio searches of \fermi\ sources \citep{ckr+12}. Surveys such as these are rapidly expanding the known population of MSPs, which may help to better understand the nature of binary and isolated MSPs. Also, continued long-term timing has shown that some of these new  MSPs are useful in pulsar timing arrays for gravitational wave detection \citep{haa+10,abb+15,2022Sci...376..521F}. 
Here we present an overview of the most prolific of the early \fermi{} targeted searches, reporting the discovery of 10 MSPs and the detection of eight more MSPs that were found in  subsequent surveys. 

This paper presents our survey results in detail \citep[a brief overview of the new discoveries was published previously in][]{hrm+11}. Here, we give detailed radio timing solutions for the ten MSPs discovered in this survey. We also present the detection of pulsed $\gamma$-ray emission from nine of these. Six of these were previously reported in the second LAT pulsar catalog \citep[][hereafter 2PC]{2PC}, and all were included in the third LAT pulsar catalog \citep[][hereafter 3PC]{3PC}. The new discoveries include one isolated MSP and six in close binaries with periods under one day. Three of these have very low mass companions ($M_c \ll 0.1\msun$), and an additional two exhibit extensive eclipses. One of the newly discovered MSPs is a binary which appears to be a chance discovery, well outside the target LAT positional error ellipse.~In \S\ref{sec:survey} we describe the survey.~In \S\ref{sec:timing} we describe the subsequent timing program and summarize the survey results as well as describe follow-up radio, X-ray, and $\gamma$-ray studies of these sources. In \S\ref{sec:discussion} we discuss the implications of the new discoveries for different pulsar populations with detailed discussions of the new black widow and redback systems.  

%---------------------------------------------------------------------------------
\section{Search observations and results}
\label{sec:survey}

Candidate sources for our radio searches were drawn from a preliminary version of the first \fermi-LAT source catalog \citep[1FGL,][]{aaa+10b}. We considered the fraction of the sky which is visible from the GBT for more than $\sim$1~hr, or equivalently Dec. $>-$40$^\circ$. We chose 350 MHz as the observation frequency due to the steep spectra of MSPs \citep{kxl+98} and the larger beam size of the GBT at low frequencies. The positional uncertainty of the LAT sources was typically less than the 35$\amin$ full-width-half-max (FWHM) beam of the 350-MHz system. We chose sources well away from the Galactic plane, i.e., $|b|$ $>$ 5$^\circ$, where sky temperature and the effects of interstellar scattering are reduced. We excluded sources that had viable Active Galactic Nuclei (AGN) counterparts \citep{aaa+10c} and sources with statistically significant variability as defined in 1FGL, as pulsars had long been recognized as a non-variable $\gamma$-ray population \citep[e.g.][however, see \citet{rgr02,mbh+13,sah+14} for discussions of $\gamma$-ray variability from pulsar \textit{wind}\ and accretion flows; and \citet{allafort13} for the discovery of a rare example of $\gamma$-ray pulsar variability]{mmct96}. We also excluded sources in the \fermi-LAT Bright Source List \citep{aaa+09c} since those had largely been surveyed with the GBT at 820~MHz by \citet{rrc+11}. There is no clear relation between $\gamma$-ray and radio pulsed flux (3PC), which likely depends to a large degree on beaming geometry, and so targeting a fainter $\gamma$-ray population would not be expected to yield a smaller (or larger!) fraction of radio pulsar counterparts.

Among the sources that passed these criteria, we generally preferred sources that had an obviously `pulsar-like' spectrum, i.e., a power-law behavior at lower energies but with significant curvature/cutoff at higher energies. However, we did not exclude sources based solely on spectra but balanced spectral desirability with considerations of visibility and observing efficiency within each scheduled observation session. By following this procedure, 49 sources were finally observed throughout 2009. The 1FGL and 3FGL source names, pointing positions, 3FGL $\gamma$-ray fluxes, variability indices,  observation dates, exposure lengths, and minimum detectable fluxes are listed in Table~\ref{srcs}, along with source classifications.

\tabletypesize{\scriptsize}
\begin{deluxetable*}{lrrcccrcccc}
\label{srcs}
\tablecolumns{11}
\tablewidth{0pc}
\tablecaption{350-MHz Observations of \fermi-LAT Sources}
\tablehead{
    \colhead{1FGL}	&  \colhead{$l$}	&   \colhead{$b$}    & \colhead{ 3FGL} & \colhead{3FGL off} & \colhead{3FGL Flux} & \colhead{Var. Idx.} &	\colhead{Date}	&   \colhead{$t_{\mathrm{obs}}$}  & \colhead{S$_{\mathrm{min}}$}	&  \colhead{Notes} \\
	&  \colhead{deg.}	&   \colhead{deg.}    &       & \colhead{deg.} & \colhead{$10^{-9}$\,ph\,cm$^{-2}$\,s$^{-1}$} &       &	 	&   \colhead{min.}  & \colhead{mJy}	&   
}
\startdata
J0008.3$+$1452	 &  107.65	 &   $-$46.70  	     & J0008.3+1456 & 0.046 & 0.41 & 34.44 &  2009-10-24	 &   22	   &   0.15   & NVSS J000825$+$145635             \\ 
J0023.5$+$0930	 &  111.53	 &   $-$52.79  	     & J0023.4+0923 & 0.082 & 1.12 & 51.37 &  2009-10-25	 &   32	   &   0.13   &  \textbf{PSR J0023$+$0923}  		        \\
J0046.8$+$5658	 &  122.27	 &   $-$5.94   	     & J0047.0+5658 & 0.077 & 1.58 & 74.44$^{\blacktriangle}$ &  2009-10-25	 &   32	   &   0.21   & GB6 J0047$+$5657  \\ 
J0103.1$+$4840	 &  124.97	 &   $-$14.15  	     & J0102.8+4840 & 0.100 & 2.19 & 55.01 &  2009-11-04	 &   32	   &   0.16   &  \textbf{PSR J0102$+$4839}                 \\
J0106.7$+$4853	 &  125.50	 &   $-$13.89  	     & J0106.5+4855 & 0.025 & 3.36 & 41.67 &  2009-10-25	 &   32	   &   0.16   &  PSR J0106$+$4855$^{\bullet}$		\\
J0226.3$+$0937	 &  158.20	 &   $-$46.63  	     & J0226.3+0941 & 0.041 & 0.68 & 95.92$^{\blacktriangle}$ &  2009-11-04	 &   32	   &   0.14   & NVSS J022634$+$093843 \\ 
J0305.0$-$0601	 &  185.36	 &   $-$51.88  	     & None & -- & -- & -- &  2009-11-04	 &   32	   &   0.10   &  PMN J0304--0608			\\
J0308.6$+$7442	 &  131.73	 &      14.23	     & J0308.0+7442 & 0.016 & 3.04 & 50.41 &  2009-10-25	 &   32	   &   0.17   &  \textbf{PSR J0307$+$7443}		\\
J0311.3$-$0922	 &  191.40	 &   $-$52.54  	     & None & -- & -- & -- &  2009-11-04	 &   32	   &   0.10   &  			\\
J0340.4$+$4130	 &  153.81	 &   $-$11.02  	     & J0340.3+4130 & 0.029 & 3.66 & 60.97 &  2009-10-25	 &   32	   &   0.16   &  \textbf{PSR J0340$+$4130} 		\\
J0523.5$-$2529	 &  228.24	 &   $-$29.80  	     & J0523.3--2528 & 0.043 & 1.77 & 50.15 &  2009-10-25	 &   18	   &   0.13   &  CRTS J052316.9$-$252737   \\ 
J0533.9$+$6758	 &  144.78	 &      18.16	     & J0534.0+6759 & 0.037 & 1.40 & 60.02 &  2009-10-24	 &   28	   &   0.17   &  PSR J0533$+$6759$^{\bigstar}$ \\
J0545.6$+$6022	 &  152.45	 &      15.74	     & J0545.6+6019 & 0.045 & 1.27 & 35.95 &  2009-10-25	 &   32	   &   0.17   &                  \\ 
J0547.0$+$0020c	 &  205.14	 &   $-$14.15  	     & J0546.4+0031c & 0.244 & 0.57 & 53.31 &  2009-10-27	 &   32	   &   0.16   &  		\\
J0622.2$+$3751	 &  175.84	 &      10.96	     & J0622.2+3747 & 0.061  & 2.18 & 53.97 &  2009-10-27	 &   32	   &   0.15   &  PSR J0622$+$3749$^{\bullet}$	\\
J0803.1$-$0339	 &  224.67	 &      14.09	     & J0803.3--0339 & 0.088  & 0.99 & 69.53 &  2009-10-27	 &   18	   &   0.16   &  TXS 0800$-$034         \\ 
J0843.4$+$6718	 &  147.70	 &      35.58	     & J0843.4+6713 & 0.085 & 0.64 & 55.97 &  2009-10-27	 &   32	   &   0.13   &  PSR J0843$+$67$^{\dagger}$           \\ 
J0902.4$+$2050	 &  206.66	 &      37.74	     & J0902.4+2050 & 0.025 & 1.23 & 101.83$^{\blacktriangle}$ &  2009-10-27	 &   32	   &   0.10   & NVSS J090226$+$205045 \\ 
J0929.0$-$3531	 &  263.05	 &      11.21	     & J0928.9--3530 & 0.030 & 0.64 & 37.60 &  2009-10-24	 &   32	   &   0.17   & NVSS J092849$-$352947   \\ 
J0953.6$-$1505	 &  251.85	 &      29.66	     & J0953.7--1510 & 0.094 & 1.25 & 40.21 &  2009-10-24	 &   32	   &   0.11   &                         \\ 
J0955.2$-$3949	 &  269.96	 &      11.49	     & J0954.8--3948 & 0.118 & 1.30 & 51.34 &  2009-10-24	 &   32	   &   0.14   &  PSR J0955$-$3947$^{\dagger}$ \\ 
J1119.9$-$2205	 &  276.51	 &      36.04	     & J1119.9--2204 & 0.038 & 2.70 & 62.62 &  2009-10-24	 &   32	   &   0.11   & CRTS J111958.3$-$220456	   \\ 
J1124.4$-$3654	 &  284.17	 &      22.79	     & J1123.9--3653 & 0.085 & 2.18 & 34.58 &  2009-10-24	 &   32	   &   0.13   &  \textbf{PSR J1124$-$3653} 		\\
J1142.7$+$0127	 &  267.51	 &      59.44	     & J1142.9+0120 & 0.033 & 1.00 & 70.84 &  2009-10-24	 &   32	   &   0.12   &  PSR J1142$+$0119$^{\bigstar}$ \\
J1302.3$-$3255	 &  305.60	 &      29.90	     & J1302.3--3259 & 0.075 & 1.99 & 39.75 &  2009-10-24	 &   32	   &   0.18   &  \textbf{PSR J1302$-$3258}		\\
J1312.6$+$0048	 &  314.73	 &      63.20	     & J1312.7+0051 & 0.066 & 2.41 & 46.35 &  2009-10-24	 &   32	   &   0.13   &  PSR J1312$+$0051$^{\bigstar}$ \\
J1544.5$-$1127	 &  356.22	 &      32.96	     & J1544.6--1125 & 0.026 & 1.01 & 47.47 &  2009-10-24	 &   32	   &   0.17   &  1RXS J154439.4$-$112820  \\ 
J1549.7$-$0659	 &  1.23	 &      35.01	     & J1549.7--0658 & 0.006 & 0.96 & 48.92 &  2009-10-24	 &   32	   &   0.15   &  \textbf{PSR J1551$-$0658}		\\
J1600.7$-$3055	 &  344.06	 &      16.46	     & J1600.8--3053 & 0.031 & 1.08 & 42.59 &  2009-11-04	 &   32	   &   0.21   &  PSR J1600$-$3053$^{\vartriangle}$ \\
J1627.6$+$3218	 &  52.99	 &      43.28	     & J1627.8+3217 & 0.074 & 0.57 & 33.03 &  2009-11-06	 &   32	   &   0.13   &  PSR J1627$+$3219$^{\dagger}$          \\ 
J1722.4$-$0421	 &  18.30	 &      17.60	     & J1722.7-0415 & 0.203 & 1.02 & 34.03 &  2009-11-04	 &   32	   &   0.22   &                          \\ 
J1730.7$-$0352	 &  19.92	 &      15.98	     & J1730.6--0357 & 0.113 & 1.33 & 29.97 &  2009-11-04	 &   32	   &   0.23   &                          \\ 
J1806.2$+$0609	 &  33.32	 &      12.83	     & J1805.9+0614 & 0.127 & 1.12 & 40.32 &  2009-11-04	 &   32	   &   0.25   &  PSR J1805$+$0615$^{\dagger}$	 \\
J1810.3$+$1741	 &  44.61	 &      16.86	     & J1810.5+1743 & 0.054 & 2.54 & 51.86 &  2009-10-25	 &   32	   &   0.20   &  \textbf{PSR J1810$+$1744} 		\\
J1858.1$-$2218	 &  13.57	 &   $-$11.40        & J1858.2--2215 & 0.047 & 1.65 & 55.58 &  2009-10-24	 &   32	   &   0.25   &  PSR J1858$-$2216$^{\bigstar}$		\\
J1903.8$-$3718c	 &  359.76	 &   $-$18.39        & None & -- & -- & -- &  2009-10-24	 &   32	   &   0.22   &   			\\
J1921.2$+$0132	 &  37.70	 &   $-$5.93         & J1921.2+0136 & 0.090 & 1.27 & 35.34 &  2009-11-03	 &   32	   &   0.36   & PSR J1921$+$01                \\ 
J2023.7$-$1141	 &  32.62	 &   $-$25.69        & J2023.6--1139 & 0.007 & 1.05 & 62.82 &  2009-10-24	 &   32	   &   0.14   & PMN J2023$-$1140              \\ 
J2043.2$+$1709	 &  61.89	 &   $-$15.32        & J2043.2+1711 & 0.021 & 4.40 & 59.09 &  2009-10-24	 &   32	   &   0.15   & PSR J2043$+$1711$^{\vartriangle}$ 		\\
J2055.2$+$3144	 &  75.35	 &   $-$8.62         & None & -- & -- & -- &  2009-10-24	 &   32	   &   0.18   &   			\\
J2057.4$+$3057	 &  75.00	 &   $-$9.43         & None & -- & -- & -- &  2009-10-25	 &   32	   &   0.18   &   			\\
J2107.5$+$5202c	 &  92.23	 &      3.07         & J2108.1+5202 & 0.117 & 1.66 & 57.21 &  2009-10-25	 &   32	   &   0.33   &                             \\ 
J2112.5$-$3044	 &  14.90	 &   $-$42.44        & J2112.5--3044 & 0.009 & 3.26 & 51.84 &  2009-10-24	 &   32	   &   0.14   &                             \\ 
J2129.8$-$0427	 &  48.97	 &   $-$36.96        & J2129.6--0427 & 0.071 & 0.91 & 60.25 &  2009-10-24	 &   32	   &   0.14   &  \textbf{PSR J2129$-$0429}		\\
J2139.9$+$4715	 &  92.62	 &   $-$4.03         & J2140.0+4715 & 0.024 & 3.86 & 39.37 &  2009-10-25	 &   32	   &   0.22   &  PSR J2139$+$4716$^{\bullet}$		\\
J2204.6$+$0442	 &  64.80	 &   $-$38.66        & J2204.4+0439 & 0.059 & 0.49 & 59.90 &  2009-11-04	 &   32	   &   0.12   &  4C $+$04.77                   \\ 
J2216.1$+$5139	 &  99.92	 &   $-$4.18         & J2215.6+5134 & 0.048 & 2.17 & 56.86 &  2009-10-25	 &   32	   &   0.19   &  \textbf{PSR J2215$+$5135} 		\\
J2256.9$-$1024	 &  59.21	 &   $-$58.29  	     & J2256.7--1022 & 0.054 & 1.32 & 33.45 &  2009-10-24	 &   32	   &   0.12   &  PSR J2256$-$1024$^{\dagger}$ 		\\
J2257.9$-$3643	 &  4.03	 &   $-$64.21        & J2258.2--3645 & 0.070 & 0.27 & 33.94 &  2009-11-04	 &   32	   &   0.12   &  MRSS 406$-$025483                           \\ 
\enddata
\tablecomments{Columns include the 1FGL name of the source, Galactic longitude $l$ and latitude $b$, the 3FGL name of the source, the offset of our pointing center from the center of the 3FGL position ellipse in degrees (note that all beams with 3FGL counterparts completely cover the 3FGL error ellipse), the 1$-$100 GeV $\gamma$-ray flux in photons~cm$^{-2}$~s$^{-1}$, the \gray~variability index (with higher values indicating greater probabilities of being a variable source, with a typical range for non-variable sources of 35--70), the date of the observation, the minimum detectable flux (in mJy), calculated from Eq.~\ref{eq:smin}, and a note indicating a source classification, if identified.  The pulsars discovered in this survey are indicated with bold face. \\
$\bigstar$: $\gamma$-ray MSP discovered in 820\,MHz survey and subsequently detected in 350\,MHz data. \\
$\vartriangle$: MSP already discovered in other pulsar survey. \\
$\bullet$: Young blind search $\gamma$-ray pulsar. \\
$\blacktriangle$: \gray~variability index above 72.44, which indicates the source is variable on the scale of months with $>99\%$ confidence. This information was not available at the time of source selection.\\
$\dagger$: Recently-discovered MSPs: \url{http://astro.phys.wvu.edu/GalacticMSPs/GalacticMSPs.txt}} 
\end{deluxetable*}

The observations were performed using the 4096-channel 100-MHz bandwidth mode of the GUPPI (Green Bank Ultimate Pulsar Processing Instrument) backend on the GBT \citep{drd+08} with a central frequency of 350~MHz. The integration times were generally 32 minutes, making this about an order of magnitude more sensitive than any prior large-scale surveys of the Northern sky and five to six times more sensitive than the Green Bank North Celestial Cap
pulsar survey \citep{slr+14}, at least for sources near the center of our beam and assuming typical pulsar spectral indices. The sensitivity of the search is obtained by using the modified radiometer equation:  

\begin{equation}
S_\mathrm{min} = \frac{(S/N)_\mathrm{min}
~(T_\mathrm{sys}+T_\mathrm{sky})}{G \sqrt{n_{p} t_\mathrm{obs} \triangle f}} \sqrt{\frac{W}{P-W}}
\label{eq:smin}
\end{equation}
\\
where \textit{(S/N)$_\mathrm{min}$} = 8 is the threshold signal to noise ratio used, \textit{G} = 2 K Jy$^{-1}$ is the effective 
gain of the GBT, \textit{n$_{p}$} = 2 is the number of polarizations summed, \textit{$\triangle f$} = 100 MHz is the total observing bandwidth, \textit{$T_\mathrm{sys}$} = 25 K is the system temperature, \textit{$T_\mathrm{sky}$} is the sky background temperature, \textit{$t_\mathrm{obs}$} is the integration time, \textit{P} is the period of the pulsar, and \textit{W} is the pulse width.
For example, given \textit{W} = 0.1\textit{P}, typical for MSPs, a sky background temperature of 50 K, typical for  sources off of the Galactic plane at this frequency \citep{hss+82}, and an integration time of 32 minutes, the minimum detectable flux is 0.2 mJy. We list the nominal sensitivity towards each pointing position in Table~\ref{srcs}, assuming  \textit{W} = 0.1\textit{P}.

Our actual sensitivity at the beam center was likely somewhat, and perhaps substantially, worse than what would be inferred from the radiometer equation. This equation does not take into account the effect of radio frequency interference (RFI), which results in parts of the bandpass and some intervals in time being unusable. It also assumes full sensitivity across the entire observing bandwidth, whereas, in reality, the sensitivity decreases at the edges of the band, resulting in a smaller effective bandwidth. The assumption of \textit{W} = 0.1\textit{P} will not be true for all pulsars. Some may have intrinsically broadened pulse profiles, and pulsars with higher dispersion measures (DMs) may experience appreciable pulse broadening due to interstellar scattering and dispersion. In addition, it is important to note that pulsars with average fluxes greater than the quoted minimum detectable flux could be missed due to interstellar scintillation, which can cause the fluxes of pulsars to vary dramatically from epoch to epoch. Sensitivity is also worse for very short orbit binaries ($P_b \lesssim 5$~hr) because we only use an acceleration search over the first-period derivative (see below), which cannot fully correct for the orbital acceleration over a 32-minute observation. And finally, many binary pulsars are in eclipse for a significant fraction of their orbit, and during this survey, we only observed each source one time and so could have missed systems that, on average, are quite bright.

We de-dispersed the data over the range from 0~pc~cm$^{-3}$ to twice the maximum DM predicted by the  NE2001 Galactic electron density model  in the direction of each source \citep{cl02}; our search was prior to the release of the newer YMW16 electron density model \cite{ymw17}. We performed acceleration searches up to $z_\mathrm{max}$ (the maximum Fourier frequency derivative) of 200 to improve sensitivity to short-orbit binaries. Up to eight harmonics were summed in the Fourier power spectra using the standard tools found in PRESTO\footnote{http://www.cv.nrao.edu/$\sim$sransom/presto/} \citep{rem02}. 

In our survey data, we discovered 10 MSPs and detected four more that had recently been discovered by other surveys. A subsequent survey using the GBT at a central frequency of 820\,MHz targeted more than 100 LAT sources \citep{2016PhDT.......539S}, several of which we had also targeted with our 350-MHz survey. These discoveries are summarized in Table~\ref{tab:pulsars}. This higher frequency survey discovered five pulsars within our sample. Subsequent folding of our data using the 820-MHz determined periods and DMs resulted in clear detections of four of these pulsars, although mostly at significance levels below our criteria for a  discovery.  Of these four, PSRs J0533+6759, J1312+0051, and J1858$-$2216 have broad pulses, detected with low signal-to-noise in our data. The other, PSR J1142+0119, was discovered first in the 820-MHz data. The final 820-MHz discovery, PSR~J1921+0137, is not detectable in our survey, likely due to either a flat spectrum, an eclipse, or scintillation. 

\begin{figure*}
\centering
\includegraphics[scale=0.6]{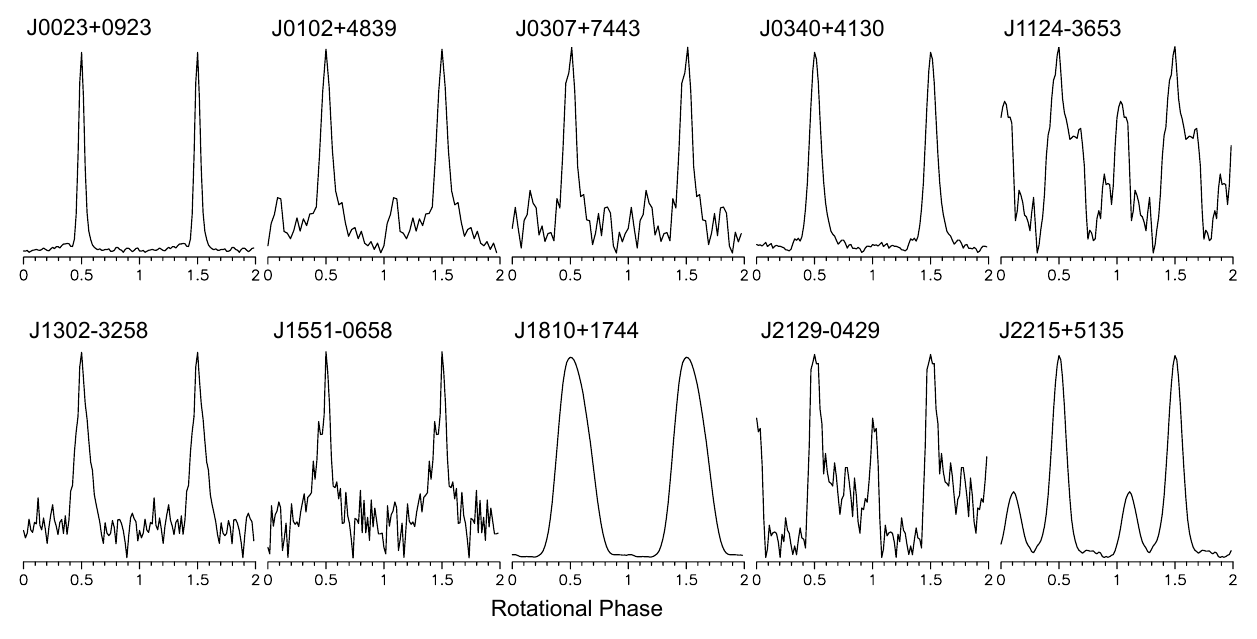}
\caption{Pulse profiles from the 350-MHz discovery observations for newly discovered MSPs. These are shifted so that the main peak is at phase 0.5. Two cycles are shown for clarity. Image credit: \cite{hrm+11}} \label{fig:profiles}
\end{figure*}

All of the 10 discoveries are MSPs, with the longest spin period being 7.6~ms. Our survey also included three LAT sources which were later determined to be young, slow  pulsars through blind searches of the
LAT $\gamma$-ray data \citep{pga+12}. Folding our data with the $\gamma$-ray period over a plausible range of DMs failed to detect any of these three.  These results are summarized in Table~\ref{tab:pulsars}, and the 350-MHz pulse profiles of the 10  discoveries are shown in Figure~\ref{fig:profiles}.
The mean flux densities in the table have been calculated using the radiometer equation (Eq. 1) and the appropriate sky temperatures for each source \citep{hss+82}. No estimates for statistical uncertainty for variability due to scintillation are available.

As described above, our pointing positions were based on an early version of 1FGL. Improved positions and $\gamma$-ray spectra were published in the second and third \fermi-LAT catalogs \citep[2FGL and 3FGL,][]{naa+12,3fgl}, and so we list the associated 3FGL source and the offset from our pointing position (in degrees) in Table~\ref{srcs}. 
In most cases, our pointing positions were consistent with the 3FGL positions, and the entire 3FGL error ellipse was contained within the half-power beamwidth. However, five of our pointings have no 3FGL counterpart within the FWHM beam, presumably due to either improved background modeling or previously confused sources being resolved. 
Our large beam also allowed for detecting pulsars well outside the LAT error ellipses, as was the case for our chance discovery of PSR J1551$-$0658. For the other 18 MSPs, the pulsars are apparently associated with the 3FGL sources. Seven 3FGL sources that we searched have now been plausibly associated with AGN; 3FGL\,J0523.5$-$2529 \citep{scs+14} and J1119.9$-$2205 \citep{Swihart22} are low-mass binary candidates; and 3FGL\,J1544.6$-$1125 is associated with a transitional MSP candidate \citep{Bogdanov15}.  We list these new associations in Table~\ref{srcs}.

In the long interval between the survey observations and the finalization of this manuscript, the 4FGL \fermi{} point source catalogs \citep{4fgl} have been released.  Because the $\gamma$-ray sources reported in this work are relatively bright, the properties are already well-determined in the 3FGL catalog, and we determined that revising all quantities to their 4FGL values is not warranted. 

%---------------------------------------------------------------------------------
\section{Radio timing observations and results}
\label{sec:timing}

Following each discovery, we began regular timing observations with the GBT  as soon as was feasible, which in a few cases was up to six months after the initial discovery observation.  All GBT observations used the GUPPI
spectrometer\footnote{https://wikio.nrao.edu/bin/view/CICADA/GUPPiUsersGuide},
typically at a central frequency of 820\,MHz due to receiver availability, and when possible at 350\,MHz. There were additional timing observations done at central frequencies of 1500 and 1900 MHz for several of the pulsars. For PSRs~J0340+3140 and J1551$-$0568, the GBT campaigns were supplemented by observations with the Nan\c{c}ay telescope in France at frequencies of  1376, 1408, and 1598 MHz. For PSR~J1551$-$0568, the position outside the \fermi\ error box was first determined with an imaging/pulsed observation by the Giant Metrewave Radio Telescope (GMRT) at a frequency of 322~MHz \citep{Roy13}. Two of our sources, PSRs~J0023+0923 and J0340+4130, have been included as targets in the NANOGrav Pulsar Timing Array and so are being extensively timed \citep{nano21a, nano21b}.  For more details of exact observations obtained for each pulsar, see the accompanying files, which contain the pulse time-of-arrival (TOA) data.

We observed each pulsar at least once a month for a span of no less than one year, with at least one densely sampled set of observations over a period of a few weeks to obtain phase-connected solutions.  Individual integration times were adjusted according to pulsar strength, and ranged from 5 minutes to 0.5\,hr.  In some instances, we performed gridding observations at 820 or 1420\,MHz to obtain immediately improved positions that increased observing efficiency and simplified the determination of timing solutions.  In a few cases, X-ray observations from a Neil Gehrels Swift Observatory (\textit{Swift}) campaign to systematically survey unassociated \fermi-LAT sources provided higher precision candidate positions \citep{fsf+11}.  Until obtaining enough timing points to yield an adequate orbital solution for each pulsar, we recorded data in search mode, which were analyzed using PRESTO \citep{rem02}. In some cases, once a reasonable initial solution was obtained, we recorded data in full-Stokes (polarimetry) fold mode, with 256 bins per pulse period. A 1-minute observation with a 25-Hz pulsed noise diode was performed before each observation for polarization calibration. These data were analyzed with \textsc{Psrchive} \citep{hvm04}. In either case we used 2048 frequency channels typically spanning 100~MHz of band at a central frequency of  350\,MHz, 200\, MHz of band at 820\, MHz, or 800\, MHz of band at the higher frequencies (1500 and 1900\,MHz).

To determine the TOAs, a high S/N profile at each frequency was used as a reference template. TOAs were then calculated by cross-correlating each pulse profile with the reference template profile at each frequency \citep{t92}.  A model ephemeris was then fitted to the topocentric TOAs.  In the case of a binary, an initial binary solution ephemeris was obtained by fitting a sinusoid to the measured barycentric frequencies. Timing solutions accompanying this paper were obtained with \textsc{Tempo2} \citep{hem06} and generally using the JPL DE421 Solar System model (Standish 2004).  Exceptions are noted below.

Six of our discoveries and two of the re-detections are in tight binaries with orbital periods $P_B < 1$\, day. Three of the short-period binaries have very low-mass companions ($M_c \ll 0.1 M_{\odot}$), and so we identify them with the `black widow' class of  MSPs. Two of these black widows, PSRs J1124$-$3653 and J1810+1744, also exhibit eclipses. Two of the other tight binaries, PSRs J2129$-$0429 and J2215+5135, have minimum companion masses in the $M_c \sim 0.2-0.4~M_{\odot}$ range, exhibit extensive eclipses, and have very strong orbital period variation. We classify them as `redbacks' \citep{rob11}. The final  short-period binary, PSR J1302$-$3258, occasionally seems to eclipse at low frequencies for~$\sim$10\% of the orbit but has a companion mass consistent with a white dwarf and shows no other characteristics of being a redback. Three of the other discoveries--PSRs J0102+4839, J0307+7443, and J1551$-$0658--are in long-period ($P_B > 1$~day) binaries around companions whose minimum masses are consistent with being white dwarfs. Only two of the detected MSPs are isolated; one of them, PSR J0533+6759, is a weak re-detection of a pulsar first discovered at 820~MHz. The other, PSR J0340+4130, is an isolated MSP which we study in this paper.

Because the LAT data span over 10 years, we found that the timing solutions established solely via these radio pulsar timing campaigns were sometimes inadequate to accurately predict the pulse phase over the full duration. This is particularly true for the compact binaries exhibiting nondeterministic orbital period variations. Consequently, we extended the timing solutions using the \fermi-LAT data.  In brief, we used maximum likelihood methods similar to those in \citet{ray11} and \citet{kerr15}.  However, these approaches are based on first estimating TOAs from LAT data.  Here, we directly optimize the unbinned likelihood using PINT\footnote{https://github.com/nanograv/PINT} \citep{luo21} and estimate parameters and their uncertainties using standard methods.  Parameters like spin frequency and its derivative and proper motions, whose effect on pulsar timing residuals is cumulative in time, are often best estimated using LAT data, while others benefit from the generally higher precision of the radio TOAs.  For each pulsar, we determined the optimal set of parameters to fit using each data set, and we iteratively fit the timing solution to the two data sets independently, yielding a timing solution that adequately describes both data sets and provides optimal parameter estimates.  For the two redback pulsars, J2129$-$0429 and J2215+5135, the orbital period variations were so substantial that we were unable to obtain a timing solution in this manner.  Instead, we used only LAT data and followed the method of \citet{clark20} to estimate the orbital period evolution.

The timing solutions\footnote{We exclude the complex orbital period variations for PSR~J2129$-$0429 and PSR~J2215+5135; see https://fermi.gsfc.nasa.gov/ssc/data/access/lat/ephems/ for these values.} for the 10 newly discovered MSPs are presented in Tables~\ref{tab:timing_msps1} and \ref{tab:timing_msps2}. For some pulsars, an ``EFAC'' term is included to describe additional scatter in the radio data beyond that indicated simply by radiometer noise. \maura{MAM: Not for all pulsars?}  Errors reported by \textsc{Tempo2} and given in these tables are scaled \maura{using this EFAC} such that the $\chi^2$ of the residuals is unity. For all of the  MSPs, in addition to the measured $\dot{P}$ we give an ``intrinsic'' $\dot{P}$ which has been corrected for acceleration due to proper motion when available (i.e., the \citet{s70} effect), acceleration toward the Galactic plane, and differential acceleration parallel to the Galactic plane \citep{nt95}. We also calculate a transverse velocity, given the DM-derived distance. Because of uncertainties in the distance and the Galactic rotation model, there is additional uncertainty in these estimates, which is hard to quantify. We use our best estimate of the intrinsic $\dot{P}$ for all derived quantities ($\dot E$, $B$, and $\tau$) in Tables~\ref{tab:timing_msps1} and \ref{tab:timing_msps2}.
The flux densities given are mean values of detections (not including eclipse times) calculated using the modified radiometer equation at 350, 820, and 1500 MHz. 

\tabletypesize{\small}
\begin{deluxetable*}{lcccccccc}
\tablewidth{0pt}\tablecolumns{9}
\tablecolumns{9}
\tablecaption{Properties of pulsars observed in this survey\label{tab:pulsars} }

\tablehead{
\colhead{Name} &\colhead{$P$} & \colhead{$S_{350}$}   & \colhead{$F_{\gamma}$\tablenotemark{\scriptsize{1}}} & \colhead{$\dot E$\tablenotemark{\scriptsize{2}}} & \colhead{DM}      &\colhead{Dist.}\tablenotemark{\scriptsize{3}}   &\colhead{$P_B$}  &\colhead{$M_c$}\tablenotemark{\scriptsize{4}} \cr
 \colhead{}
               &\colhead{ms}  &\colhead{mJy} & \colhead{$10^{-11}{\rm erg}\,{\rm cm}^{-2}{\rm s}^{-1}$}  & \colhead{$10^{34}\,$ergs\,${\rm s}^{-1}$} & \colhead{pc cm$^{-3}$} &\colhead{kpc} &\colhead{days}  &\colhead{$M_{\odot}$}}
\startdata
\cutinhead{Pulsars discovered in searches of 350~MHz data}
J0023+0923      & 3.05    & 5.4   & $0.7 \pm 0.1$    & 1.3     & 14.3   & 1.2     & 0.14      & 0.01 \\
J0102+4839      & 2.96    & 2.8   & $1.6 \pm 0.1$    & 1.6	   & 53.5   & 2.3     & 1.6       & 0.18 \\
J0307+7443      & 3.16    & 1.8   & $1.6\pm 0.1$     & 2.2     & 6.3    & 0.4     & 37.1      & 0.20  \\
J0340+4130      & 3.30    & 6.5   & $2.1 \pm 0.1$    & 0.7     & 49.6   & 1.6     & Isolated  & N/A \\
J1124$-$3653    & 2.41    & 2.2   & $1.3 \pm 0.1$    & 1.5	   & 44.9   & 1.0     & 0.23      & 0.02 \\
J1302$-$3258    & 3.77    & 1.3   & $1.2 \pm 0.1$    & 0.4     & 26.2   & 1.4     & 0.78      & 0.15\\ 
J1551$-$0658\tablenotemark{\scriptsize{5}} & 7.09    & 0.7     & --    & 0.2	  & 21.6      & 1.3     & 5.2    & 0.20 \\ 
J1810+1744      & 1.66    & 78    & $2.3 \pm 0.1$    & 3.1     & 39.6   & 2.4     & 0.15      & 0.04 \\
J2129$-$0429    & 7.61    & 2.1   & $0.8 \pm 0.1$    & 2.9     & 16.9   & 1.4     & 0.64      & 0.37 \\
J2215+5135      & 2.61    & 16    & $1.3 \pm 0.1$    & 6.1 	   & 69.2   & 2.8     & 0.17      & 0.21 \\
\cutinhead{Pulsars detected in searches of 350~MHz data, discovered previously or concurrently} 
J1600$-$3053    & 3.60    & 2.7   & $0.6 \pm 0.1$    & 0.7     & 52.3   & 2.5     & 14.3      & 0.20 \\
J1805+0615      & 2.13    & 1.1   & $0.6\pm0.1$      & 7.3      & 64.9   & 3.8     & 0.34      & 0.02  \\
J2043+1711      & 2.38    & 0.8   & $3.0\pm 0.1$      & 1.2     & 20.7   & 1.5     & 1.4       & 0.17 \\
J2256$-$1024    & 2.29    & 8.3   & $0.8 \pm 0.1$    & 3.4     & 13.8   & 1.3     & 0.21      & 0.03 \\
\cutinhead{Pulsars discovered in 820-MHz survey detected by folding 350~MHz data} 
J0533+6759      & 4.39    & 2.1   & $1.0\pm0.1$      & 0.5*    & 57.4   & 2.4     & Isolated & N/A \\
J1142+0119      & 5.07    & 0.4   & $0.6\pm0.1$      & 8.5    & 19.2   & 2.2     & 1.6   & 0.15 \\
J1312+0051      & 4.23    & 0.4   & $1.7\pm0.1$      & 0.4    & 15.3   & 1.4     & 38.5  & 0.18 \\
J1858$-$2216    & 2.38    & 2.0   & $0.8\pm0.1$      & 1.1*   & 26.6   & 0.9     & 46.1  & 0.21 \\
\cutinhead{Pulsars discovered in 820-MHz survey not detected in 350MHz data} 
J0843+67        & 2.84    & --    &                  & --      & 20.7   & 1.5    & 7.4   & 0.27 \\
J1921+0137      & 2.50    & --    & $1.6\pm0.2$      & 4.9*   & 104.9  & 5.0     & 9.9   & 0.23 \\
\cutinhead{\fermi-LAT pulsars observed but not detected in 350~MHz data} 
J0106+4855      & 83.16   & $< 0.15$  & $1.9\pm0.1$  & 2.9*   & 70.9 & 3.0 & -- & -- \\
J0622+3749      & 333.21  & $< 0.15$  & $2.0\pm0.1$  & 2.7*   & -- & -- & -- & -- \\
J1627+3219      & 2.18    & --        &              & --     & 28.1 &  & 0.18 & 0.02  \\
J2139+4716      & 282.85  & $< 0.22$  & $2.3\pm0.2$  & 0.3*   & -- & -- & -- & -- \\
\enddata
\tablenotetext{1}{Gamma-ray fluxes in the first group are from this work.  Others are taken from the 3FGL catalog.}
\tablenotetext{2}{The $\dot{E}$ values have been corrected for the Shklovskii \citep{s70} effect (except for those marked by * ) when proper motion information is available, and all have been corrected for the perpendicular and parallel components of acceleration with respect to the Galactic plane \citep{nt95}.}
\tablenotetext{3}{The distance is estimated using the YMW16 electron distribution model \citep{ymw17}. The distances reported in \citet{hrm+11} were based on the NE2001 model, as YMW16 model was not complete then.}
\tablenotetext{4}{The minimum companion masses ($M_c$) were calculated assuming a pulsar mass of 1.4 M$_{\odot}$ and an orbital inclination of 90$^{\circ}$.}
\tablenotetext{5}{Not associated with the LAT source.}
\end{deluxetable*}

%---------------------------------------------------------------------------------
\subsection{Radio polarization observations and results}

For some of the discoveries, full-Stokes polarization profiles were obtained using the coherent fold mode of the GBT GUPPI backend, and the data were analyzed using \textsc{Psrchive} \citep{hvm04}. We performed  RFI excision in frequency and in time, and the data were calibrated by ensuring the noise power was the same in both polarizations and combined.  We fit for rotation measures (RMs) where possible, and include them as RM-corrected polarization profiles in our multi-frequency profiles in Figures~\ref{fig:gamrad_j0102}--\ref{fig:gamrad_j0340}. Two of the pulsars show high degrees of linear polarization ($\sim$70\% in the main peak for PSR~J0340+4130 and nearly 100\% in the secondary peak for PSR~J0102+4839). PSR~J0307+7443's profile shows significant, $\sim$20\% linear polarization. The main peak of PSR~1551$-$0658's 820-MHz profile is $\sim$40\% linearly polarized. None of the pulsars show high degrees of circular polarization. Position angles (P.A.s) are displayed for bins surpassing a S/N threshold of 5. For PSR~J0102+4839, the position angle shows an orthogonal jump at the peak of the main pulse, suggesting another emission mode at the emission peak \citep{jvh+14}. For PSRs~J0340+4130 and J1551$-$0658, the P.A. swing is smooth across the profile, suggesting the emission originates from a continuous region.~The P.A. could not be measured for PSR~J0307+7443. 

It is sometimes possible to estimate the magnetic inclination angle $\alpha$ and the observer's viewing angle with respect to the rotation axis $\zeta$ from the rotating-vector model \citep{rc69}. However, MSP polarization profiles often contain features---like orthogonal jumps---which cannot be adequately described by the model.  In none of our data sets is a direct constraint on $\alpha$ or $\zeta$ available.  Rough constraints on $|\alpha-\zeta|<\approx30^\circ$ are consistent with visible radio beams.

\begin{figure}
\begin{centering}
\includegraphics[scale=0.75,angle=0,width=0.98\linewidth]{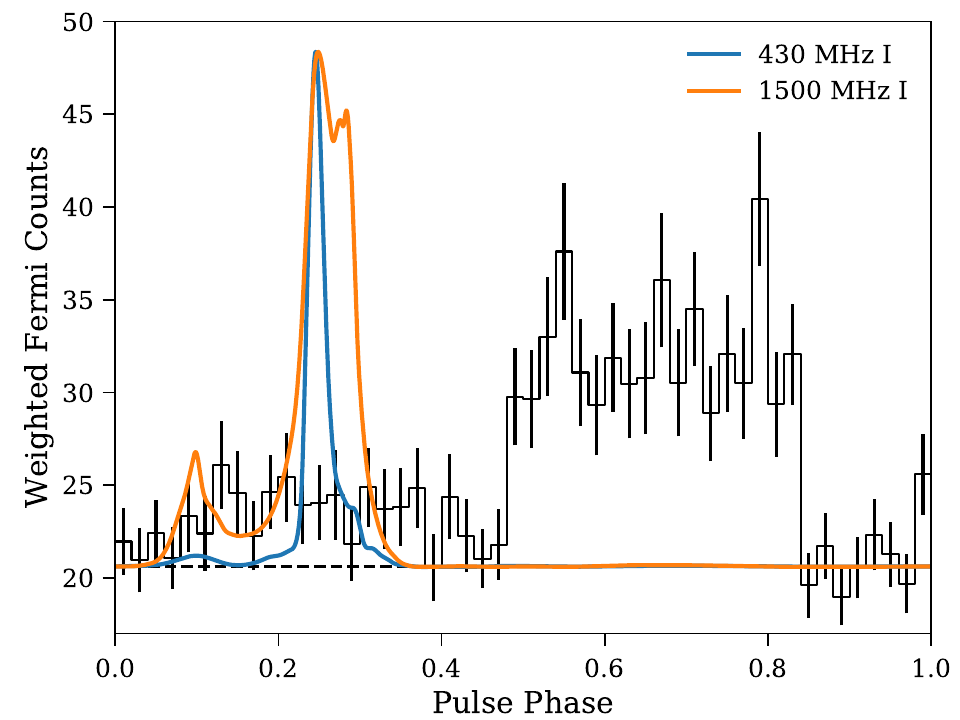}
\caption{Phase-aligned weighted $\gamma$-ray (black) and 430 and 1500-MHz radio polarimetric (blue and orange) pulse profiles for PSR~J0023+0923 (from Arecibo NANOGrav observations). The $\gamma$-ray pulse profile indicates the pedestal of background emission as estimated by spectral analysis. Note the clear precursor pulse and double peaked main pulse at 1500~MHz as compared to the 430~MHz pulse here and the 350~MHz pulse shown in Figure~\ref{fig:profiles}. By 2~GHz, the second peak of the main pulse dominates (see Figure 22.2 in 2PC), shifting the apparent phase by $\sim 0.05$ as compared to the 350~MHz main peak. The radio and $\gamma$-ray peaks are misaligned with the main radio peak at pulse phase $\phi \sim 0.25$ preceding the broad $\gamma$-ray peak centered $\sim 0.65$. \label{fig:gamrad_j0023} }
\end{centering}
\end{figure}

\begin{figure}
\begin{centering}
\includegraphics[scale=0.75,angle=0,width=0.98\linewidth]{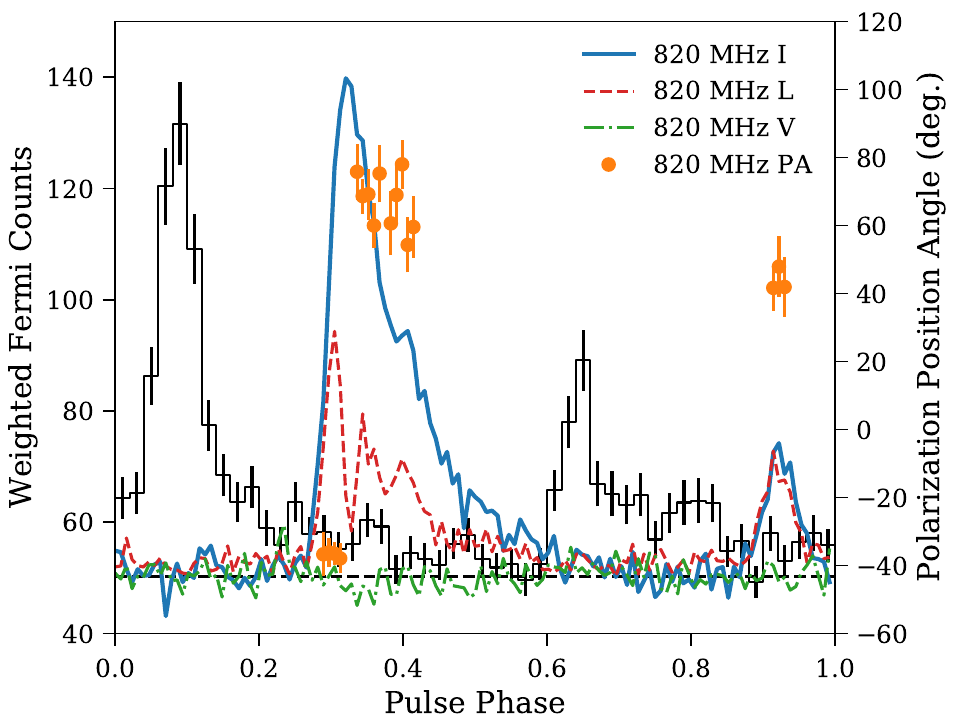}
\caption{Phase-aligned $\gamma$-ray (black) and 820-MHz radio polarimetric (blue) pulse profiles for PSR~J0102+4839. In the radio profile, linear (circular) polarization is shown in red (green), while the polarization position angle appears in orange. The radio and $\gamma$-ray peaks are misaligned. The main radio peak at $\phi \sim 0.35$ is followed by at least one additional component, a ``shoulder'' at $\phi \sim 0.45$. We do not believe this is due to scatter-broadening as this feature is less prominent at lower frequencies. There is a second peak at $\phi \sim 0.90$; it is not clear whether we are seeing emission from one or both magnetic poles. Note the orthogonal mode jump at the peak of the main pulse. \label{fig:gamrad_j0102} }
\end{centering}
\label{fig:0102}
\end{figure}

\begin{figure}
\begin{centering}
\includegraphics[scale=0.75,angle=0,width=0.98\linewidth]{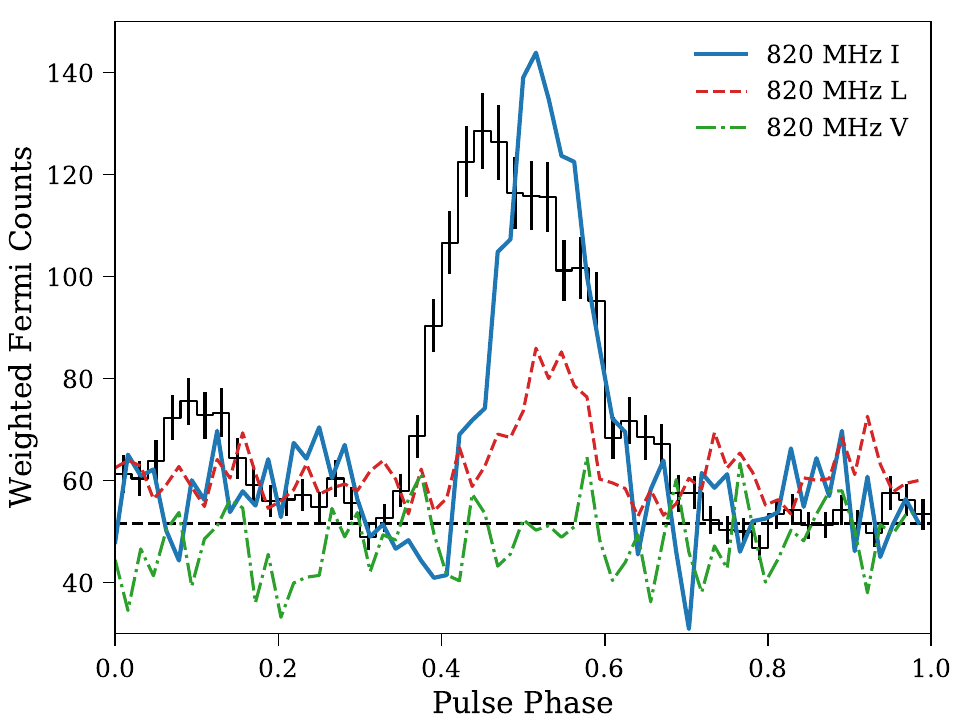}
\caption{\label{fig:gamrad_j0307}
Phase-aligned $\gamma$-ray (black) and radio (blue) pulse profiles for PSR~J0307+7443. The main $\gamma$-ray and radio peaks are roughly aligned. In the radio profile, linear (circular) polarization is shown in red (green). Due to the low signal-to-noise ratio, a position angle measurement cannot be made.}
\end{centering}
\label{fig:0307}
\end{figure}

\begin{figure}
\begin{centering}
\includegraphics[scale=0.75,angle=0,width=0.98\linewidth]{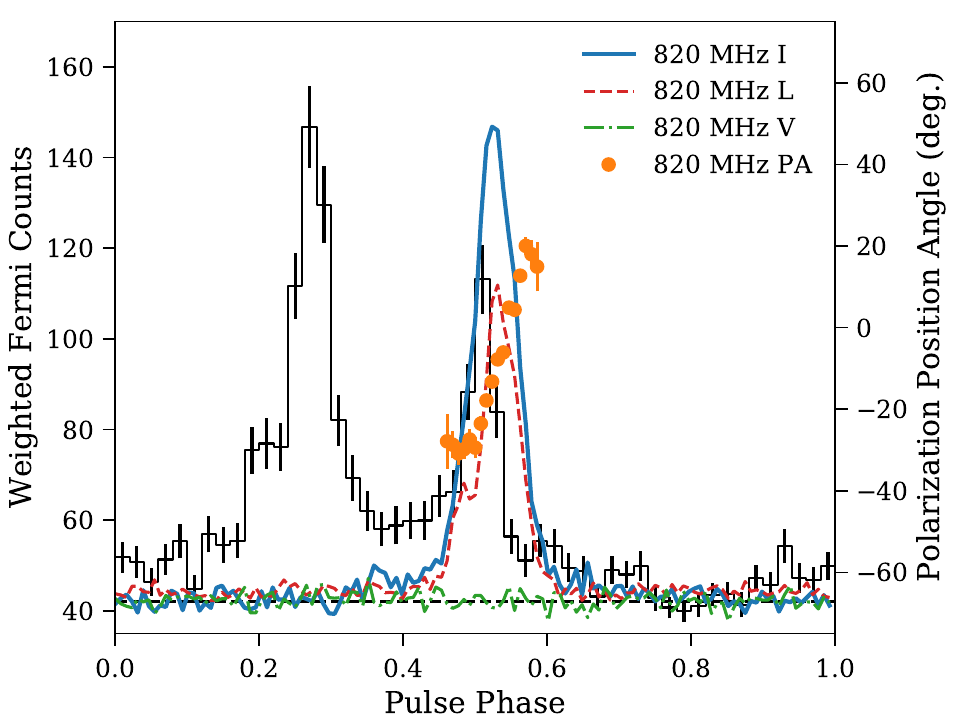}
\caption{\label{fig:gamrad_j0340}
Phase-aligned $\gamma$-ray (black) and 820-MHz radio polarimetric (blue) pulse profiles for PSR~J0340$+$4130. In the radio profile, linear (circular) polarization is shown in red (green), while the polarization position angle appears in orange. The radio peak at $\phi\simeq0.55$ is aligned with one of the $\gamma$-ray peaks.}

\end{centering}
\label{fig:0340}
\end{figure}

%---------------------------------------------------------------------------------
\subsection{X-Ray observations and analysis}

\citet{grm+14} presented \textit{Chandra} observations covering a full orbit of the three new black widows from this survey (PSRs J0023+0923, J1124$-$3553, and J1810+1744) as well as the redback PSR~J2215+5135. They found clear indications of orbitally varying hard X-ray emission from PSR J2215+5125 (see below for an \textit{XMM-Newton} analysis) and tentative evidence for such variations in PSR~J1124$-$3653. All four of these systems also had a soft X-ray component consistent with the thermal emission seen from other MSPs. The remaining redback from our survey, PSR J2129$-$0429, was observed by \textit{XMM-Newton} %%(PI M. Roberts) 
over an entire orbit. Preliminary results were presented in \citet{rmg+15} \citep[see also][]{hhp+15}, and a more detailed analysis, including \textit{NuSTAR} hard X-ray observations, in \citet{alnoori18}. It has a predominantly non-thermal spectrum with a flat power-law component extending out to at least 40 keV. The flux of this hard component varies by more than a factor of 10 over the orbit.

\begin{figure}
\begin{centering}
\hspace*{-0.6cm}
\includegraphics[scale=1.0,angle=0,width=1.1\linewidth]{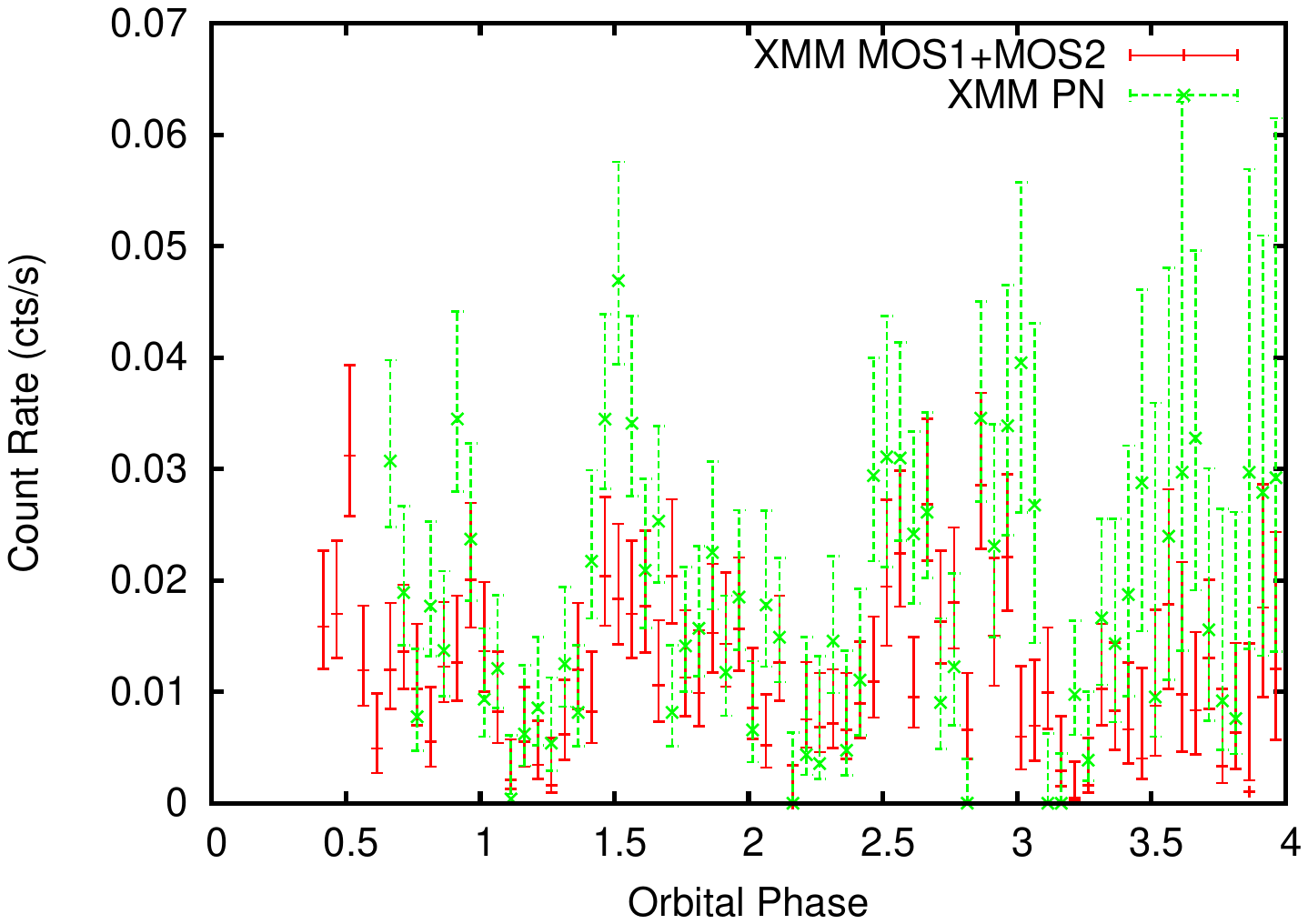}
\caption{\xmm~background-subtracted 0.5-7 keV light curve of PSR J2215+5135, produced from the full observation, which extends for about 3.5 orbits.  Orbital phase is defined so that pulsar superior junction is at 0.25. Green indicates the PN data, and red indicates the summed MOS1 and MOS 2 data. The error bars represent the 68\% confidence region of the Bayesian posterior probability distribution of the count rate using an exposure corrected background region. \label{fig:2215_xmm_lc} }
\end{centering}
\end{figure}

\begin{figure}
\begin{centering}
\hspace*{-0.6cm}
\includegraphics[scale=1.0,angle=0,width=1.1\linewidth]{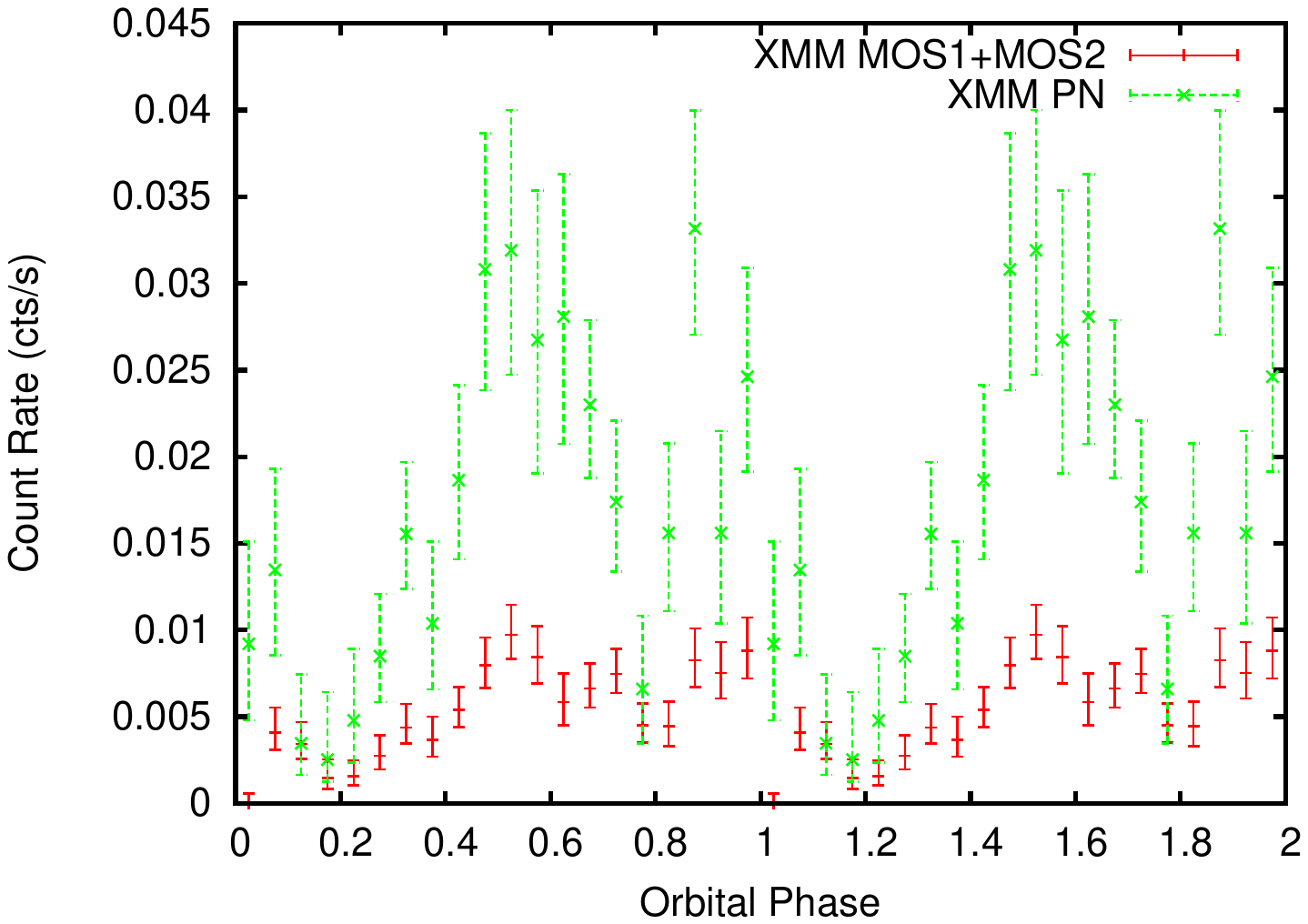}
\caption{Background-subtracted \xmm~0.5-7.5 keV light curve folded on the orbital period, repeated twice for clarity. \label{fig:2215_xmm_fold} }
\end{centering}
\end{figure}

\xmm~observed PSR J2215+5125 on 2016-06-17 for 54~ks (around  3.5 orbits). For all three imaging detectors, we extracted and barycentered 0.5-7 keV events from a $20^{\prime \prime} $ radius region centered on the timing position and from a nearby $40^{\prime \prime} $ radius background region. We constructed binned background subtracted source intensity light curves and 68\% confidence regions using the Bayesian method outlined by \citet{l92} for the summed light curves of the MOS 1 and 2 detectors and for the PN detector shown in Figure~\ref{fig:2215_xmm_lc}. We also produced a light curve folded on the orbital period using the same method and show that light curve in Figure~\ref{fig:2215_xmm_fold}. 

The \xmm~data was spectrally fit to an absorbed blackbody plus power law model over the $0.2-10$ keV band.  The hydrogen column density ($n_H$) 
was fixed at $2.41 \times 10^{21} {\tt cm^{-2}}$ derived from an $A_V$ calculated from the Milky Way dust model found in \cite{gsf+15}. We fit   a flux $F_{bb}\sim 10^{-14}~ {\rm erg~ cm^{-2}~s^{-1}}$, $kT=0.26\pm0.05$\, keV to the thermal component, typical for the surface of a neutron star \citep{Marelli11}. The power-law component is very hard, with a photon index of $\Gamma = 1.0\pm 0.2$ and a flux $F_{pl}\sim 10^{-13}\,  {\rm erg~ cm^{-2}~s^{-1}} $, similar to other known redbacks. A more complete analysis of these data will be presented elsewhere. 

Here, we also report on X-ray observations of the other five discoveries from this survey.  Because in none of these cases a detailed spectral analysis possible, we first describe our general approach and assumptions.  To estimate expected X-ray absorption, we used the model of optical extinction of the Galaxy by \citet{dcl03} to estimate the visual extinction $A_V$ towards the source at the nominal DM-derived distance, and then the empirical relationship of $N_H \sim 2.21\times 10^{21} A_V~{\rm cm^{-2}}$ of \citet{go09}. To estimate the expected emission, we assumed a blackbody of $kT = 0.2$~keV, which is typical for the majority of MSPs which do not have intrabinary shock emission or strong pulsed X-ray emission. %\footnote{See online list and references maintained by P. Gentile at {\tt http://astro.phys.wvu.edu/XrayMSPs/xraymsps.html}.}.
Flux limits quoted are for the 0.3--8~keV range. Note that assuming a power law of 1.5 (reasonable for magnetospheric emission or intrabinary shock emission) would increase the flux limit by a factor of $\sim$2--3.

PSR~J0102+4839 is one of many sources which has benefited from a \textit{Swift} campaign to monitor \textit{Fermi} sources \cite{fsf+11} with the XRT \citep{bhn+00}.  Observations in October and November 2010 resulted in a total exposure of 4412~s. No excess emission was seen at the pulsar position, placing a conservative upper limit of $10^{-14}\,{\rm erg~cm^{-2}~s^{-1}}$ on the 0.3--8 keV flux. %%emission.

PSR J0307+7443 was observed by XRT three times in 2009--2010 for a total exposure of 10458~s. No clear detection was made, although there is a local $\sim 1.5 \sigma$ excess at the pulsar position, which would correspond to a 0.3--8~keV flux of $\sim$$10^{-14}\,{\rm erg~cm^{-2}~s^{-1}}$. 

{\it XMM-Newton} observed PSR J0340+4130 twice in August 2009 for a total of 41~ks. However, the observations were affected by background flaring, and after excising flares, there was a total exposure kept for analysis of 21.6, 25.9, and 11.6~ks for the MOS1, MOS2, and PN detectors, respectively.  No significant emission was found at the position of the pulsar. Our best estimate of the count rate from the PN is $0.0007\pm 0.0006$ cts/s. From this we derive an upper limit of $\sim$2$\times 10^{-15} {\rm erg~cm^{-2}~s^{-1}}$. 

{\it Chandra} observed PSR J1302$-$3258 for 9.9~ks on 30 March 2011 with the {\it ACIS-S} instrument. A total of 16 counts were detected in the 0.3--2.0 keV range within $2^{\prime \prime}$ of the pulsar position, with an estimated background of $\sim$0.1 counts. No photons above 2~keV were detected, suggesting the emission to be predominantly thermal emission from the pulsar surface. Using an estimated neutron hydrogen column density $n_H=5\times 10^{20} {\rm cm^{-2}}$, this corresponds to a 0.2~keV blackbody flux of $1.2\times 10^{-14}\,{\rm erg~cm^{-2}~s^{-1}}$ or an unabsorbed flux of $1.6\times 10^{-14}{\rm erg~cm^{-2}~s^{-1}}$, yielding an estimated isotropic X-ray luminosity of $1.9\times 10^{30}\,d_{\mathrm{1kpc}}^2$~erg~s$^{-1}$. 

PSR J1551$-$0658, being a chance detection, is $21.7^{\prime}$ from the nearest {\it Swift} pointing of the nearby {\it Fermi} source, and was hence out of the field of view. We therefore have no X-ray data on the source. 

Given the DM-derived distance estimates, these luminosities and limits are all consistent with what would be expected from the rough relationship between spin-down power and total X-ray emission derived for MSPs with reliable parallax derived distances, ${\rm log} L_x = (0.27\pm 0.20){\rm log} \dot E + 20.90\pm6.53$ \citep{bognar}. They are also in line with the relation for just the blackbody emission component, ${\rm log} L_x = (0.25\pm 0.16){\rm log} \dot E + 21.28\pm5.36$, which tends to dominate, although that of PSR J1302$-$3258 is towards the high end of the scatter in the data. 

To summarize, while one of the black widows and both redbacks show clear evidence for non-thermal emission,  there is no evidence for either bright magnetospheric or intrabinary shock emission from any of the other pulsars, with all the data being consistent with the thermally dominated pulsar surface emission typically seen from MSPs. 

%---------------------------------------------------------------------------------
\subsection{$\gamma$-ray observations and analysis}

To characterize the $\gamma$-ray emission and pulsations, we analyzed $\sim$8 years (2008 August 4 to 2016 July 20) of P8R2 \fermi-LAT data.  All LAT data analyses were carried out with v10r00p05 of the \fermi\ Science Tools (STs)\footnote{http://fermi.gsfc.nasa.gov/ssc/data/analysis/software/}.  We selected \texttt{SOURCE} class events, as defined under the \texttt{P8R2\_V6} instrument response functions (IRFs), which had energies between 100 MeV and 100 GeV, reconstructed directions within 15\degree\ of the radio position of each MSP, and zenith angles $\leq$ 90\degree.  We used the ST \texttt{gtmktime} to exclude times when the LAT was not in nominal science operations mode, when the LAT data were flagged as bad, and during bright LAT-detected solar flares and gamma-ray bursts.

We performed a binned maximum-likelihood analysis on a $21\fdg2\times21\fdg2$ region for each MSP using the \texttt{P8R2\_SOURCE\_V6} IRFs, combining events which convert in the front and back section of the LAT.  We calculated binned exposure maps with 30 logarithmic energy bins and spatial pixels that were $0\fdg1$ on a side.  The source model for each region included all 3FGL point and extended sources within $25^{\circ}$ of each MSP and the diffuse background models.  The Galactic diffuse emission was modeled using the \textit{gll\_iem\_v06.fits} template, and the isotropic diffuse spectrum, which includes the extragalactic diffuse and residual instrument backgrounds, was modeled using \textit{iso\_P8R2\_SOURCE\_V6\_v06.txt}.  The spectral parameters of point sources within 6\degree\ of each MSP and with average significance $\geq10\sigma$, in four years, were left free as well as the normalizations of the Galactic and isotropic diffuse components and point sources within 8\degree\ that were flagged as a variable in 3FGL.  All other spectral parameters were kept fixed to the 3FGL values.

Since we are using twice as much (8 years vs.~4 years) and improved (P8 vs.~P7REP) data compared to the 3FGL Catalog, it is possible that our models of the regions may be incomplete. After our initial fits, we looked at the spatial and spectral residuals to see if additional sources needed free parameters or if additional sources needed to be included.  For the region around PSR J2215+5135, we did find a few sources outside the 6\degree\ radius that were over or undersubtracted. We refit this region with those additional source normalizations free and found satisfactory fits.  In the region around PSR J0340+4130, we saw evidence for a source not included in 3FGL.  Using the ST \texttt{gtfindsrc}, we localized this source to (R.A., Dec.) = ($59\fdg88$, $50\fdg96$) with a 95\% confidence-level error radius of $0\fdg03$, positionally associating this source with the flat-spectrum radio quasar, 4C $+$50.11, which flared in $\gamma$ rays in 2014 January \citep{4CATel}.  After including it in the model, we found satisfactory fits.

All of the new  MSPs, with the exception of PSR~J1551$-$0658, were found as significant point sources in the 3FGL catalog, with positions consistent with the timing positions within the LAT error ellipses.  We fixed them to their timing positions in the region models and modeled their spectra as a power law (Equation~\ref{PLeqn}) and an exponentially-cutoff power law (Equation~\ref{COeqn} with $b\ \equiv\ 1$).

\begin{equation}\label{PLeqn}
\frac{dN}{dE} = K \left( \frac{E}{E_{0}} \right)^{-\Gamma}
\end{equation}
\begin{equation}\label{COeqn}
\frac{dN}{dE} = K \left( \frac{E}{E_{0}} \right)^{-\Gamma} \exp{\left( - \frac{E}{E_{\mathrm{C}}} \right)^b}
\label{eq:expcutoff}
\end{equation}

\noindent{}For both spectral models $K$ is a normalization factor, with units of cm$^{-2}$ s$^{-1}$ MeV$^{-1}$; $E_{0}$ is a scale energy, fixed to the 3FGL pivot energy value for each MSP; and $\Gamma$ is the photon index.  For the cutoff model, $E_{\mathrm{C}}$ is the cutoff energy and $b$ is an exponential index controlling the sharpness of the cutoff.  We define a test statistic (TS) as twice the difference in the log likelihood to assess the overall source significance and to test nested models.  The power-law model is rejected (large TS$_{\mathrm{cutoff}}$) in favor of the cutoff model with $>$ 5.5 $\sigma$ significance for all detected MSPs.  We also performed analysis with the $b$ parameter free but found no significant improvement (small TS$_{\mathrm{b\,free}}$, $\lesssim$2.5$\sigma$) of the fit for any MSP when doing so.  The best-fit spectral parameters and TS values for all nine MSPs are given in Tables~\ref{tab:timing_msps1} and \ref{tab:timing_msps2}.

We used the best-fit models and \texttt{gtsrcprob} to calculate spectral weights (probability that each event came from the respective MSP based on the likelihood model) for each event.  We then folded the events for each MSP using the ephemerides described in \S\ref{sec:timing} using the \texttt{fermi} plugin to \textsc{tempo2} \citep{ray11} and calculated a weighted H-test \citep{Kerr11} for each MSP.  We formed $\gamma$-ray profiles by generating weighted histograms from events lying within 3\degrees{} of the source and display them in Figures 2--5, and FIgures 8--12.  Estimates of the background level (dashed line) and error flags for the intensity in a given phase bin are computed via a bootstrap process described in 2PC.  In each panel, we show one or more radio pulse profiles to place the $\gamma$-ray emission in context.

\tabletypesize{\scriptsize}
\begin{deluxetable*}{lccccccccccc}
\tablewidth{0pt}
\tablecaption{Timing and $\gamma$-ray Spectral Parameters for the Discovered MSPs\label{msp_timing1}}
%--------------------------------------------------------------------------------------------------------------------------------------------------------------------------------------------------------------------------------
\tablehead{\colhead{Timing Parameters} 		& \colhead{PSR J0023$+$0923} 	& \colhead{PSR J0102$+$4839} 	& \colhead{PSR J0307$+$7443} 	& \colhead{PSR J0340$+$4130} 	& \colhead{PSR J1124$-$3653} }
\startdata
\fermi-LAT 3-Year Source \dotfill 	& 3FGL\ J0023.4$+$0923		& 3FGL\ J0102.8$+$4840 		& 3FGL\ J0308.0$+$7442 		& 3FGL\ J0340.3$+$4130 		& 3FGL\ J1123.9$-$3653 	\\
%--------------------------------------------------------------------------------------------------------------------------------------------------------------------------------------------------------------------------------
Right Ascension (J2000)  \dotfill 	& $00^{\rm h}\;23^{\rm m}\;16\fs88027(1)$    & $01^{\rm h}\;02^{\rm m}\;50\fs66874(6)$    & $03^{\rm h}\;07^{\rm m}\;55\fs886(1)$ 	& $03^{\rm h}\;40^{\rm m}\;23\fs288229(6)$	 & $11^{\rm h}\;24^{\rm m}\;01\fs1160(2)$  	\\
Declination     (J2000)  \dotfill  	& $+09\degrees\;23\amin\;23\farcs8902(5)$     & +48$\degrees\;39\amin\;42\farcs7635(5)$     & $+74\degrees\;43\amin\;13\farcs426(5)$ 	& $+41\degrees\;30\amin\;45\farcs2892(1)$	& $-36\degrees\;53\amin\;19\farcs087(4)$ 	\\
Pulsar Period (ms) 	\dotfill   	                                & 3.05020310344768(3) 	    & 2.9641124215887(4)    	    & 3.1560886884993(7) 	        & 3.29933936625009(2)	& 2.4095726733227(3)	   \\
Period Derivative, {\it \.{P}} (10$^{-20}$ s s$^{-1}$) \dotfill & 1.14238(3)                & 1.1406(3)                     & 1.7279(7)                  	& 0.70485(8)	        & 0.60151(1)                \\
Reference Epoch (MJD) 	\dotfill 	                            & 55521		                & 55527		                    & 55780 	                    & 56279		            & 55128	                    \\
Dispersion Measure (pc cm$^{-3}$) \dotfill                      & 14.326553(4)	            & 53.5041(6)			        & 6.3430(8)			            & 49.57583(3)			& 44.8560(2)   	            \\
Proper motion in RA $\mu_\alpha\cos\delta$ (mas~yr$^{-1}$) \dotfill     	& $-$11.00(7)   & $-$3.4(4)  			        & $-$1(1)			            & $-$0.77(5)     		& 3.1(4)                    \\
Proper motion in Dec $\mu_\delta$  (mas~yr$^{-1}$) \dotfill  	& $-$8.8(1) 			    & $-$1.9(5) 			        & 3(1)			                & $-$3.1(1) 			& $-$4.1(6)	                \\
Orbital Period (days) 		 	\dotfill 	     	            & 0.13879914308(1) 		    & 1.672149564(1) 		        & 37.10775488(9) 	            & \dots				    & 0.226987946(8)     	    \\
Projected Semi-Major Axis (lt-s)  	\dotfill 	                & 0.03484228(5) 			& 1.8558825(7) 			        & 16.108338(8) 			        & \dots				    & 0.079631(6)               \\
Orbital Eccentricity  			\dotfill 	     	            & $<$3.4 $\times$ 10$^{-5}$	& $<$2.3 $\times$ 10$^{-6}$	    & $<$3.3 $\times$ 10$^{-3}$ 	& \dots				    & 0.0  \\
Epoch of Ascending Node (MJD) 		\dotfill 	                & 55186.1134208(1)		    & 55514.5773299(1)		        & 55599.363937(4) 		        & \dots				    & 55128.586598(3)           \\
Span of Timing Data (MJD) 		\dotfill 	     	            & 55130.08$-$57375.95		& 54682.78$-$59050.55 		    & 54682.77$-$59033.74		    & 55186.08$-$57378.23	& 55128.59$-$56854.65       \\
%---------------------------------------------------------------------------------------------------------------------------------------------------------------------------------------------------------------------------------
\cutinhead{Timing-Derived Parameters} 
Mass Function ($10^{-3}$ \msun) 	\dotfill 	                & 2.35$\times$10$^{-3}$		& 2.45				& 3.25				& \dots		    & 0.0105		     \\
Minimum Companion Mass (\msun) 		\dotfill 	                & $\geq$\,0.01			    & $\geq$\,0.18		& $\geq$\,0.20 		& \dots		    & $\geq$\,0.028      \\
Galactic Longitude (deg) 		\dotfill 	                    & 111.3				        & 124.7 			& 131.7	 			& 153.7		    & 284.1              \\
Galactic Latitude (deg) 		\dotfill 	                    & $-$52.8				    & $-$14.2			& 14.2 		 		& $-$11.1 		& 22.8               \\
%DM-derived Distance (NE2001, kpc) 		\dotfill 	            & 0.7				        & 2.3				& 0.6				& 1.7			& 1.7                \\
DM-derived Distance (YMW16, kpc) 		\dotfill 	            & 1.2				        & 2.3				& 0.4				& 1.6			& 1.0                \\
Surface Mag. Field, $B$ ($10^8$\,G) 	\dotfill                & 1.8			            & 1.8				& 2.3 				& 1.5			& 1.2                \\
Characteristic Age (Gyr) 		\dotfill 	                    & 4.2			            & 4.1				& 2.9	 			& 7.4			& 6.3                \\
Spin-down Luminosity, $\dot{E}$ ($10^{34}$\,erg\,s$^{-1}$)  \dotfill       & 1.58922(4)		& 1.7291(5) 		& 2.1699(9)			& 0.77477(8)	& 1.6974(5)          \\
%----------------------------------------------------------------------------------------------------------------------------------------------------------------------------------------------------------------------------
Transverse velocity (km~s$^{-1}$) \dotfill 					                & 80(16) 	    & 43(12) 	        & 6(2)	            & 25(5)		    &  25(6)   \\
Acceleration $\perp$ to plane (10$^{-14}$ s s$^{-1}$) \dotfill 	            & $-$5.8(4) 	& $-$4.9(3) 	    & 2.1(3)            & $-$4.0(2)	    & 4.4(2)    \\
Acceleration $\parallel$ to plane  (10$^{-14}$ s s$^{-1}$) \dotfill  	    & $-$0.7(1) 	& $-$0.73(2)	    & $-$0.071(8)	    & 2.1(4)	    & $-$2.0(4) \\
Corrected {\it \.{P}} (10$^{-20}$ s s$^{-1}$) \dotfill 	                    & 0.91(4) 	    & 1.096(9) 	        & 1.718(2) 	        & 0.65(8)	    & 0.55(8)   \\
Corrected $\dot{E}$ ($10^{34}$\,erg\,s$^{-1}$) 	\dotfill 	                & 1.27(6)	 	& 1.66(1)	        & 2.158(3)		    & 0.72(1)       &  1.57(2)  \\
Corrected $B$ ($10^8$\,G) 	\dotfill 		                                & 1.7	 	    & 1.8		        & 2.3		        & 1.5		    & 1.2       \\ 
%---------------------------------------------------------------------------------------------------------------------------------------------------------------------------------------------------------------------------------
\cutinhead{Observed Parameters}
Rotation Measure (RM) (rad m$^{-2}$)	\dotfill	     	    & \dots 	& $-$86.3(8)	& 13(3) 	& 56(1) 	& \dots     \\
Flux Density at 350\,MHz (mJy)          \dotfill             	& 5.4		& 2.8           & 1.8       & 6.5       & 2.2       \\
Flux Density at 820\,MHz (mJy)          \dotfill             	& 2.4		& 0.8           & 0.19      & 1.6       & \dots     \\
Flux Density at 1400\,MHz (mJy)         \dotfill             	& 0.5 		& \dots         & \dots     & 0.5       & \dots     \\
Flux Density at 2000\,MHz (mJy)         \dotfill             	& 0.5 		& \dots         & \dots     & \dots     &  \dots 	\\
%---------------------------------------------------------------------------------------------------------------------------------------------------------------------------------------------------------------------------------
\cutinhead{Spectral Fit Parameters}
K ($10^{-12}$ cm$^{-2}$\,s$^{-1}$\,MeV$^{-1}$) 	 \dotfill  	    & 3.0(5)     & 2.4(2)  		& 3.6(6) 	& 2.9(2) 	& 1.8(2)	    \\
Photon Index $\Gamma$ 				 \dotfill  	                & 1.3(2)     & 1.7(1) 		& 1.0(1) 	& 1.2(1) 	& 1.4(1)	    \\
E$_{\mathrm{C}}$ (GeV)  			 \dotfill  	                & 1.9(5)     & 5(1)			& 1.6(2) 	& 3.6(4) 	& 3.7(7)	    \\
F$_{100}$ ($10^{-8}$ cm$^{-2}$\,s$^{-1}$) 	 \dotfill  	        & 0.9(2)     & 2.0(2) 		& 1.5(1) 	& 1.5(1) 	& 1.2(1)	    \\
G$_{100}$ ($10^{-11}$ erg\,cm$^{-2}$\,s$^{-1}$)  \dotfill  	    & 0.7(1)     & 1.6(1) 		& 1.6(1) 	& 2.1(1)	& 1.3(1)	    \\
TS 						 \dotfill 	                            & 470 		 & 1349 		& 1920 		& 2448		& 1149	        \\
TS$_{\mathrm{cutoff}}$ 				 \dotfill 	                & 48		 & 66.5 		& 254 		& 200		& 85	        \\
TS$_{\mathrm{b\ free}}$ 			 \dotfill 	                & 0.9 		 & 0.0 			& 0.2 		& 0.8		& 0.5	        \\
Efficiency                           \dotfill                   & 0.10      & 0.62         & 0.013     & 0.90      & 0.10           \\
\enddata
\tablecomments{Numbers in parentheses represent 2-$\sigma$ uncertainties in the last digit as determined by {\tt TEMPO2}.  
Minimum companion masses were calculated assuming a pulsar mass of 1.4\,\msun. The gamma-ray spectral parameters are from fits of a power-law with exponential cutoff shape, 
given in Equation~\ref{eq:expcutoff} with $b$=1.  The first three parameters are as defined in Equation~\ref{eq:expcutoff}: F$_{100}$ and G$_{100}$ give the integrated photon or 
energy flux above 0.1\,GeV, respectively, while the last two parameters are gamma-ray detection significance of the source and significance of an exponential cutoff
(as compared to a simple power law).  Gamma-ray efficiency is estimated as $4\pi G_{100} d^2/\dot{E}$ using YMW16 \citep{ymw17} distances assuming spatially uniform emission.  The uncertainty on DM produces a negligible $<$0.0P uncertainty in radio/$\gamma$ alignment at 820\,MHz, with a maximum of $<$0.02P at 350\,MHz for an uncertainty of 0.001 DM units.}
\label{tab:timing_msps1}
\end{deluxetable*}

\tabletypesize{\scriptsize}
\begin{deluxetable*}{lccccc}
\tablewidth{0pt}
\tablecaption{Timing and $\gamma$-ray Spectral Parameters for the Discovered MSPs\label{msp_timing2}}

%--------------------------------------------------------------------------------------------------------------------------------------------------------------------------------------------------------------------------------
\tablehead{\colhead{Timing Parameters} 		& \colhead{PSR J1302$-$3258}	& \colhead{PSR J1551$-$0658} & \colhead{PSR J1810$+$1744} & \colhead{PSR J2129$-$0429} & \colhead{PSR J2215$+$5135} }
\startdata
\fermi-LAT 3-Year Source \dotfill 	& 3FGL\ J1302.3$-$3259 		& (3FGL\ J1549.7$-$0658) & 3FGL\ J1810.5$+$1743 & 3FGL\ 
J2129.6$-$0427 & 3FGL\ J2215.6$+$5134\\
%---------------------------------------------------------------------------------------------------------------------------------------------------------------------------------------------------------------------------
Right Ascension (J2000)  \dotfill 	& $13^{\rm h}\;02^{\rm m}\;25\fs5262(2)$	& $15^{\rm h}\;51^{\rm m}\;09\fs5279(2)$	&  $18^{\rm h}\;10^{\rm m}\;37\fs28478(5)$ & $21^{\rm h}\;29^{\rm m}\;45\fs05(8)$	& $22^{\rm h}\;15^{\rm m}\;32\fs6(2)$	\\
Declination     (J2000)  \dotfill  	& $-32\degrees\;58\amin\;36\farcs843(7)$   	&  $-06\degrees\;58\amin\;07\farcs83(1)$   	&  $+17\degrees\;44\amin\;37\farcs380(1)$  &  $-04\degrees\;29\amin\;06\farcs81(8)$	&  $+51\degrees\;35\amin\;36\farcs3(3)$   	\\
Pulsar Period (ms) 	\dotfill   	                                    & 3.770853091420(1)	            & 7.09375570886(1)	            & 1.6627549909837(1)	        & 7.613937470415(5)     & 2.6096197252726(2)        \\
Period Derivative, {\it \.{P}} (10$^{-20}$ s s$^{-1}$)  \dotfill    & 0.656(1)     	                & 2.05(6)            	        & 0.4476(1) 	                & 32.775(5)             & 2.8204(3)	                \\
Reference Epoch (MJD) 	\dotfill 			                        & 55925		                    & 55997		                    & 55530                         & 55750	                & 56362		                \\
Dispersion Measure (pc cm$^{-3}$) \dotfill                          & 26.179(1)                     & 21.578(1)	                    & 39.65952(6)	                & 16.8771(1) 	        & 69.1947(1)		        \\
Proper motion in RA $\mu_\alpha\cos\delta$ (mas~yr$^{-1}$) \dotfill     	    & \dots		 		& \dots		 		            & 7.5(2)	                    & 12.3(1)	            & 0.3(5)	                \\
Proper motion in Dec $\mu_\delta$  (mas~yr$^{-1}$) \dotfill  		& \dots		 		            & \dots		 		            &$-$3.6(4)	                    & 10.1(1)		        & 1.8(6)	                \\
Orbital Period (days) 			\dotfill 	     	                & 0.78444160(2)                 & 5.20657070(3)		            & 0.148170285(1)	            &  0.63522773(2)	    & 0.172501860(1) 		    \\
Projected Semi-Major Axis (lt-s)  	\dotfill 	     	            & 0.92792(4)                    & 4.331018(6)		            & 0.095378(1)		            & 1.85219(2)	        & 0.468131(4)			    \\
Orbital Eccentricity  			\dotfill 	     		            & 3.2 $\times$ 10$^{-5}$        & 1.1 $\times$ 10$^{-5}$ 	    & 4.7 $\times$ 10$^{-5}$        & 0.0	                & 0.0                       \\
Epoch of Ascending Node (MJD) 		\dotfill 	                    & 55924.803600(5)               & 55999.906196(1) 		        & 55529.9597206(2)	            & 55702.111161(7)	    & 55702.111161(7)		    \\
Span of Timing Data (MJD) 		\dotfill 	     	                & 54683.52$-$59050.18           & 55623.28$-$56371.38 		    & 54682.96$-$59050.59           & 54682.84$-$58900.40   & 54682.84$-$58900.40		\\
%--------------------------------------------------------------------------------------------------------------------------------------------------------------------------------------------------------------------------
\cutinhead{Timing-Derived Parameters}
Mass Function ($10^{-3}$ \msun) 	\dotfill 	                    & 1.39	   	            & 3.21	            & 0.042	            & 16.9              & 3.70			        \\
Minimum Companion Mass (\msun) 		\dotfill 	                    & $\geq$\,0.15          & $\geq$\,0.20      & $\geq$\,0.04      & $\geq$\,0.37	    & $\geq$\,0.21		    \\
Galactic Longitude (deg) 		\dotfill 	                        & 305.5                 & 1.5       		& 44.6	            & 48.8              & 99.8			        \\
Galactic Latitude (deg) 		\dotfill 	                        & 29.9                	& 34.8       	    & 16.9	            & $-$36.8 	        & $-$4.1			    \\
%DM-derived Distance (NE2001, kpc) 		\dotfill 	                & 1.0                  	& 1.1 		    	& 2.0		        & 0.9	            & 3.0			        \\
DM-derived Distance (YMW16, kpc) 		\dotfill 	                & 1.4                  	& 1.3 		    	& 2.4		        & 1.4	            & 2.8			        \\
Surface Mag. Field, $B$ ($10^8$\,G) 	\dotfill                    & 1.6                   & 3.9	      		& 0.8               & 15.9              & 2.7			        \\
Characteristic Age (Gyr) 		\dotfill 	                        & 9.0                  	& 5.5   	        & 5.8 	            & 0.36              & 1.4			        \\
Spin-down Luminosity, $\dot{E}$ ($10^{34}$\,erg\,s$^{-1}$)  \dotfill     & 0.483(1)     	& 0.227(7)	        & 3.8443(9)		    & 2.9314(5)         & 6.2654(8)			     \\
%----------------------------------------------------------------------------------------------------------------------------------------------------------------------------------------------------------------------------
Transverse velocity (km~s$^{-1}$) \dotfill 					                & \dots		    & \dots			    & 96(19)	        & 106(21)            & 25(9)		         \\
Acceleration $\perp$ to plane (10$^{-14}$ s s$^{-1}$) \dotfill 		        & 5.2(3)	    & 5.3(3)		    & 5.2(3)            & $-$5.5(3)         & $-$3.3(3)		         \\
Acceleration $\parallel$ to plane  (10$^{-14}$ s s$^{-1}$) \dotfill  		& $-$1.2(3)	    & 2.5(6)  		    & $-$1.5(7)         & $-$0.5(1)	        & $-$4.9(6)		         \\
Corrected {\it \.{P}} (10$^{-20}$ s s$^{-1}$) \dotfill 	                    & 0.608(6)		& 1.92(4)	        & 0.36(1)           & 32.2(1)  	        & 2.770(8)			     \\
Corrected $\dot{E}$ ($10^{34}$\,erg\,s$^{-1}$) 	\dotfill 	                & 0.448(5)      & 0.213(4)	        & 3.1(1)	        & 2.86(1)	        & 6.15(1)			     \\
Corrected $B$ ($10^8$\,G) 	\dotfill 		                                & 1.5	 	    & 3.7	  	        & 0.8	            & 15.8    	        & 2.7			          \\ 
%SEE NOTE ON 2215'S PROPER MOTION AT THE BOTTOM OF THE FILE
%---------------------------------------------------------------------------------------------------------------------------------------------------------------------------------------------------------------------------------
\cutinhead{Observed Parameters}
Rotation Measure (RM) (rad m$^{-2}$)	\dotfill	     	& \dots         & \dots     & \dots	    & \dots		& \dots	    \\
Flux Density at 350\,MHz (mJy)          \dotfill            & 1.3           & 0.7       & 78		& 2.1	    & 16	    \\
Flux Density at 820\,MHz (mJy)          \dotfill            & 2.4		    & 0.3       & 7 		& \dots	    & 0.8	    \\
Flux Density at 1400\,MHz (mJy)         \dotfill            & \dots         & \dots     & 1.3	    & \dots	    & 0.14	    \\
Flux Density at 2000\,MHz (mJy)         \dotfill            & \dots         & \dots     & 0.4		& \dots	    & 0.11	    \\
%---------------------------------------------------------------------------------------------------------------------------------------------------------------------------------------------------------------------------------
\cutinhead{Spectral Fit Parameters}
K ($10^{-12}$ cm$^{-2}$\,s$^{-1}$\,MeV$^{-1}$) 	 \dotfill  	    & 1.8(2)	& \dots		& 1.1(6)	& 1.6(2)	& 1.7(2)     \\
Photon Index $\Gamma$ 				 \dotfill  	                & 1.2(2) 	& \dots		& 2.0(1)	& 1.7(1)	& 1.3(2)     \\
E$_{\mathrm{C}}$ (GeV)  			 \dotfill  	                & 3.1(6)	& \dots		& 3.8(7)	& 4(1)	    & 4.2(8)     \\
F$_{100}$ ($10^{-8}$ cm$^{-2}$\,s$^{-1}$) 	 \dotfill  	        & 0.9(1) 	& $<$ 0.5	& 4.1(2)	& 1.0(1)	& 1.0(1)     \\
G$_{100}$ ($10^{-11}$ erg\,cm$^{-2}$\,s$^{-1}$)  \dotfill  	    & 1.2(1)	& \dots		& 2.3(1)	& 0.8(1)	& 1.3(1)     \\
TS 						 \dotfill 	                            & 1132 		& 4		    & 1958		& 541		& 887               \\
TS$_{\mathrm{cutoff}}$ 				 \dotfill 	                & 101 		& \dots		& 72	    & 31		& 83        \\
TS$_{\mathrm{b\ free}}$ 			 \dotfill 	                & 0.2	    & \dots		& 0.8		& 0.1		& 6.2       \\
Efficiency                           \dotfill                   & 0.64     & \dots     & 0.49     & 0.064     & 0.19 \\
%---------------------------------------------------------------------------------------------------------------------------------------------------------------------------------------------------------------------------------
\enddata
\tablecomments{Numbers in parentheses represent 2-$\sigma$ uncertainties in the last digit as determined by {\tt TEMPO2}.  
%The red lines in the pulse profile correspond to the 820\,MHz discovery pulse profiles from the GBT.  The blue lines in the pulse profiles are the {\it Fermi}\ pulse profiles.  
Minimum companion masses were calculated assuming a pulsar mass of 1.4\,\msun. The gamma-ray spectral parameters are from fits of a power-law with exponential cutoff shape, 
given in Equation~\ref{eq:expcutoff} with $b$=1.  The first three parameters are as defined in Equation~\ref{eq:expcutoff}: F$_{100}$ and G$_{100}$ give the integrated photon or 
energy flux above 0.1\,GeV, respectively, while the last two parameters are gamma-ray detection significance of the source and significance of an exponential cutoff
(as compared to a simple power law).   Gamma-ray efficiency is estimated as $4\pi G_{100} d^2/\dot{E}$ using YMW16 distances assuming spatially uniform emission.  The uncertainty on DM produces a negligible $<$0.01P uncertainty in radio/$\gamma$ alignment at 820\,MHz, with a maximum of $<$0.02P at 350\,MHz for an uncertainty of 0.001 DM units.}
\label{tab:timing_msps2}
\end{deluxetable*}
%%\end{document}
%NOTE ABOUT 2215'S PROPER MOTION:
%There's likely something going on with either 2215's PM values, the DM derived distance, or, likely, a combination of the two.
%The problem is that when you use the best fit PM and distance values to correct for the Shklovskii effect, the corrected Pdot is negative.
%If you assume the PM values are correct and use the fact that Pdot should be >0 to put an upper limit on the distance, you'll find that 
%said upper limit puts 2215 closer to us than any other source. The PM values are preferred by the F-test, though, so in lieu of a rigorous
%explanation, we just quote the velocities and forego correcting the inferred parameters.

\begin{figure}
\begin{centering}
\includegraphics[scale=0.75,angle=0,width=0.98\linewidth]{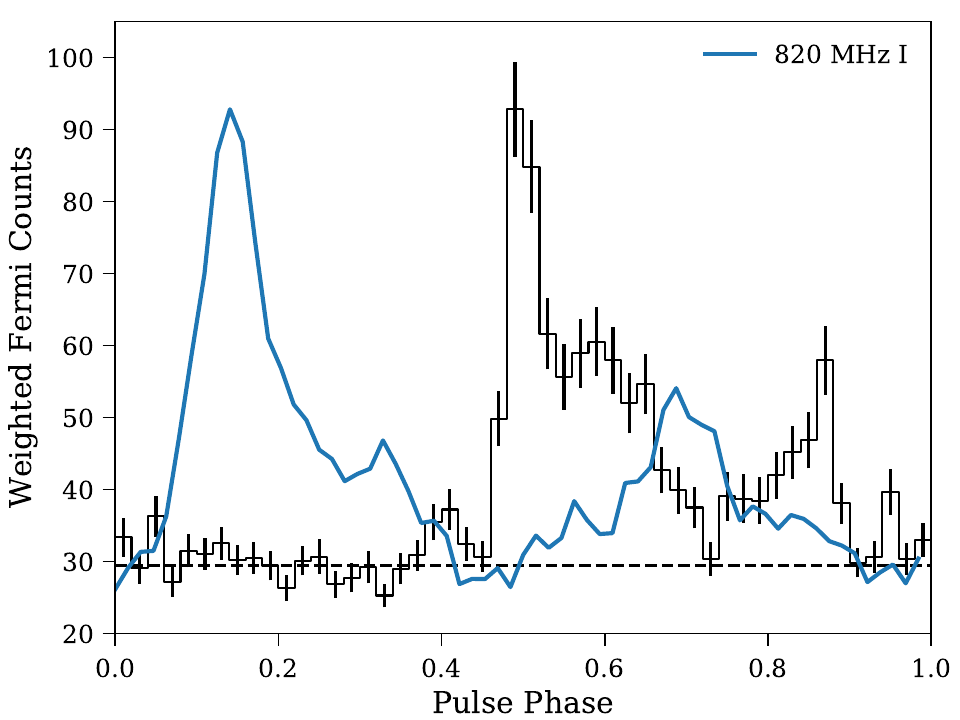}
\caption{\label{fig:gamrad_j1124}
Phase-aligned $\gamma$-ray (black) and radio (blue) pulse profiles for PSR~J1124$-$3653.  
The radio peak leads the first $\gamma$-ray peak by about 0.35\, cycles. 
}
\end{centering}
\label{fig:1124}
\end{figure}

\begin{figure}
\begin{centering}
\includegraphics[scale=0.75,angle=0,width=0.98\linewidth]{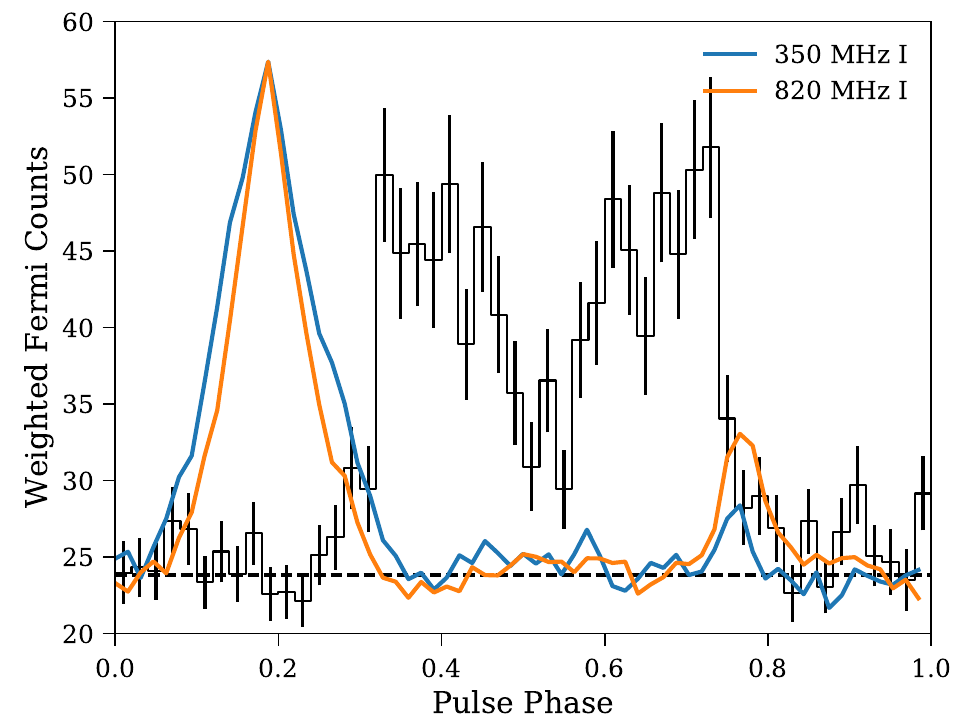}
\caption{\label{fig:gamrad_j1302}
Phase-aligned $\gamma$-ray (black) and radio (blue and orange) pulse profiles for PSR~J1302$-$3258.  
The $\gamma$-ray and radio peaks are misaligned, with a radio interpulse becoming clearly detectable at 820 MHz. }
\end{centering}
\label{fig:1302}
\end{figure}

\begin{figure}
\begin{centering}
\includegraphics[scale=0.75,angle=0,width=0.98\linewidth]{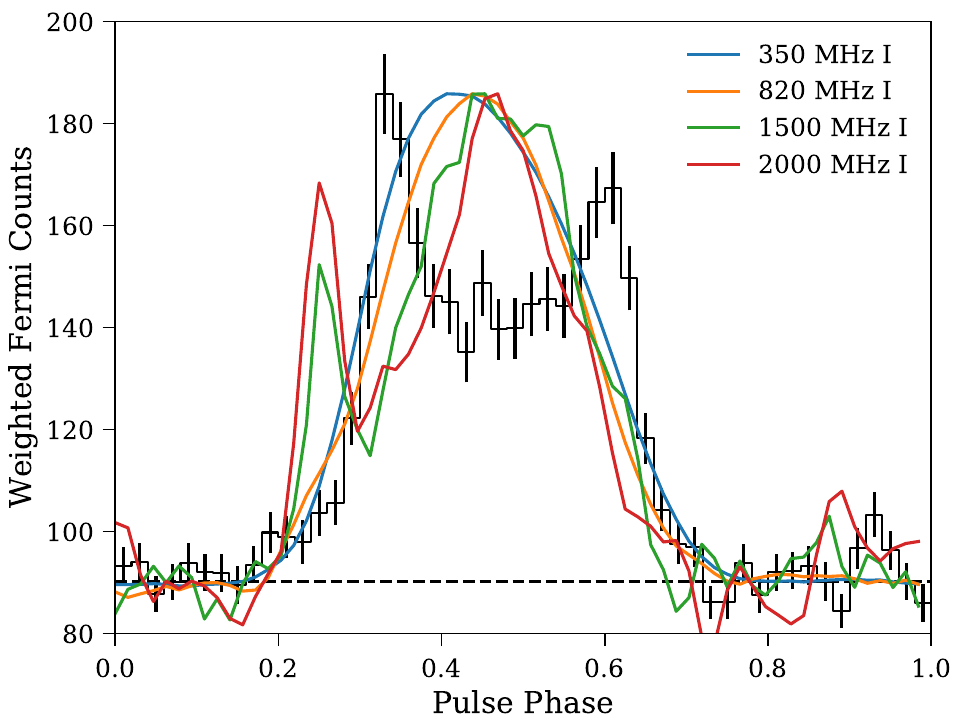}
\caption{\label{fig:gamrad_j1810}
Phase-aligned $\gamma$-ray (black) and radio (blue, orange, green, and red) pulse profiles for PSR~J1810+1744. The $\gamma$-ray and radio peaks are roughly aligned, with a radio precursor becoming clearly detectable at 1500 and 2000 MHz.}
\end{centering}
\label{fig:1810}
\end{figure}

\begin{figure}
\begin{centering}
\includegraphics[scale=0.75,angle=0,width=0.98\linewidth]{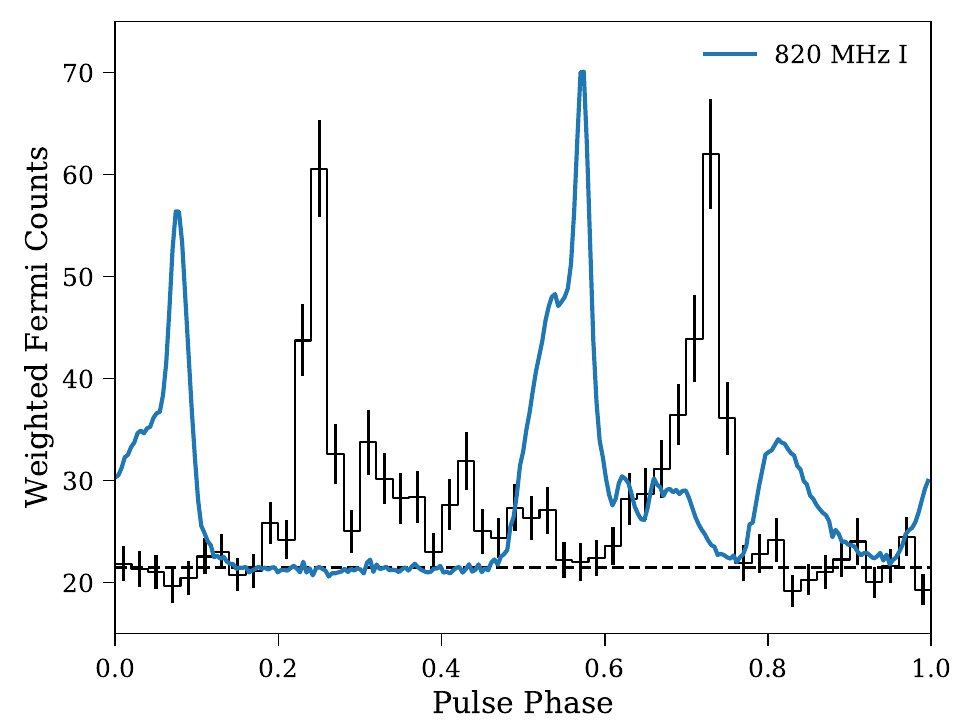}
\caption{\label{fig:gamrad_j2129}
Phase-aligned $\gamma$-ray (black) and radio (blue) pulse profiles for PSR~J2129$-$0429. The radio and $\gamma$-ray peaks are misaligned.}

\end{centering}
\label{fig:2129}
\end{figure}

\begin{figure}
\begin{centering}
\includegraphics[scale=0.75,angle=0,width=0.98\linewidth]{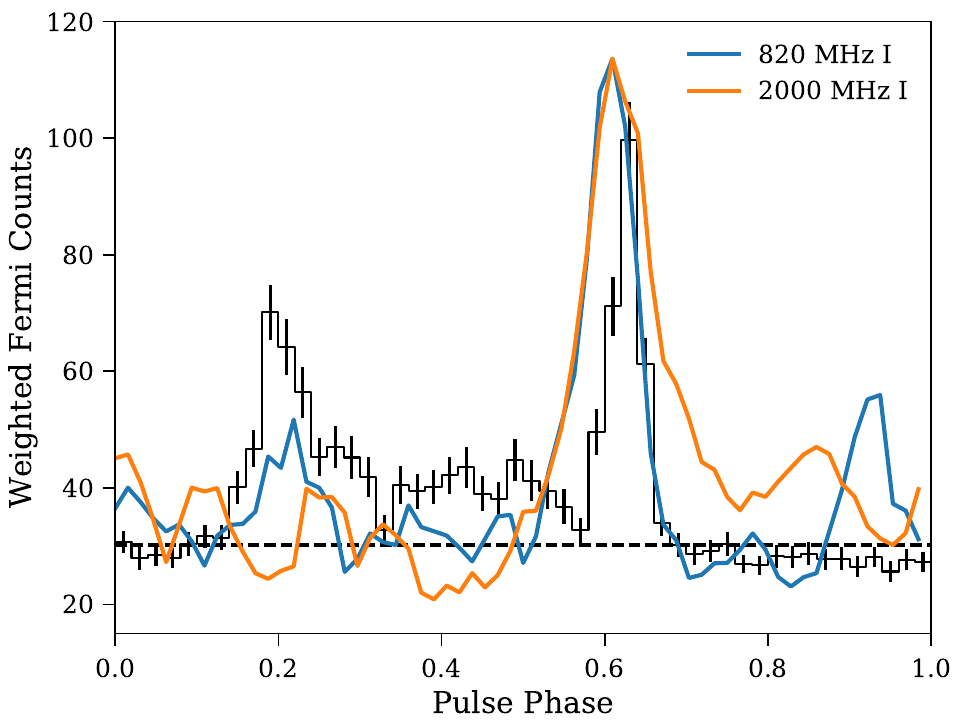}
\caption{\label{fig:gamrad_j2215}
Phase-aligned $\gamma$-ray (black) and radio (blue and orange) pulse profiles for PSR~J2215+5135. The radio peak is aligned with one of the $\gamma$-ray peaks. }

\end{centering}
\label{fig:2215}
\end{figure}

\begin{figure}
\begin{centering}
\includegraphics[scale=0.75,angle=0,width=0.98\linewidth]{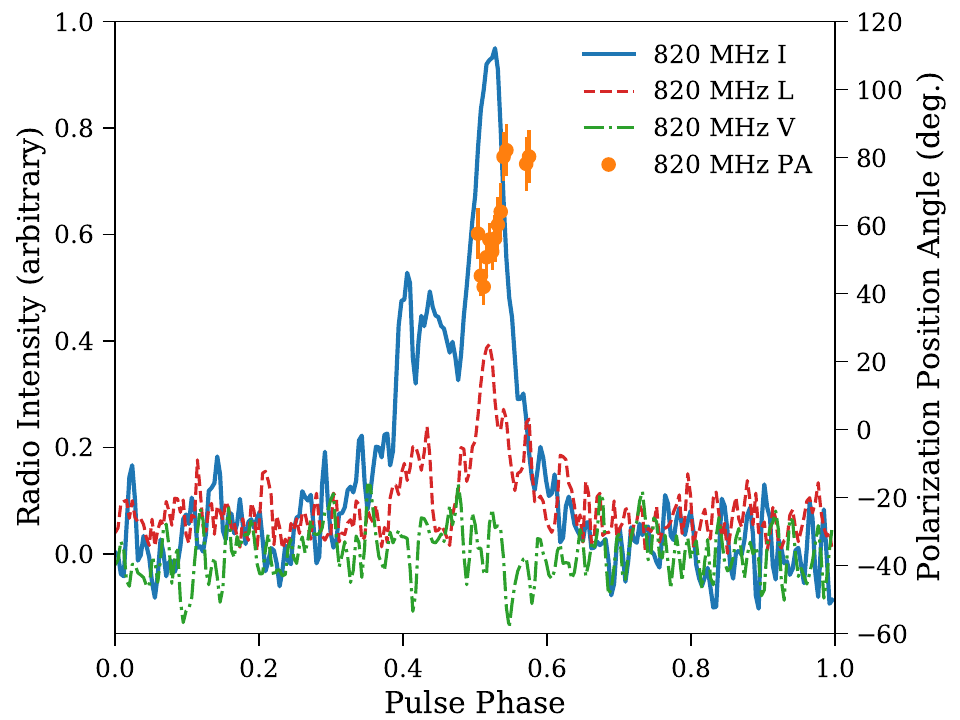} 
\caption{\label{fig:rad_j1551} 
Radio pulse profile (blue) for PSR~J1551$-$0658 at 820 MHz. The linear (circular) polarization is shown in red (green), while the polarization position angle appears in orange.}
\end{centering}
\label{fig:1551}
\end{figure}

%---------------------------------------------------------------------------------
\section{Discussion}
\label{sec:discussion}

This paper presents a 350-MHz GBT survey for pulsars, carried out in 2009, which targeted 49 unidentified \fermi-LAT sources. We detected 18 MSPs in our survey;  10 of these were discoveries unique to our survey. PSR J0340$+$4130 is an isolated MSP. PSRs J0023$+$0923, J1124$-$3653, and J1810$+$1744 are short orbital period binaries with very low mass companions ($M_c \ll 0.1M_{\odot}$), with the latter two showing eclipses at frequencies $\lesssim$2\,GHz, and hence can be classified as black widow systems. PSRs J2129$-$0429 and J2215$+$5135 have heavier, non-degenerate \citep{breton2013, bellm2013} companions of masses $M_c > 0.2 M_{\odot}$, are eclipsed for a large fraction of their orbits, and have orbitally modulated X-ray emission \citep{grm+14,rmg+15}. Moreover, this X-ray modulation is centered on pulsar inferior conjunction \citep{Wadiasingh17}.  These are hallmarks of ``redback'' systems \citep{rob11}.

PSR J1302$-$3258 also has a short binary period and shows evidence for brief ($\sim10$\% of the orbit) eclipses at low frequencies, but does not have excess X-ray emission or a bright optical companion. PSRs J0102$+$4839, J0307+7443, and J1551$-$0658 are longer period ($P_B > 1$~day) binaries with minimum companion masses typical of helium white dwarf stars orbiting MSPs. This final discovery, PSR J1551$-$0658, seems to be a chance detection since its position is well outside of the \fermi-LAT error box, and there is no evidence of pulsed emission in the $\gamma$-ray data.  We show its radio pulse profile and polarimetry in Figure \ref{fig:rad_j1551}.

%---------------------------------------------------------------------------------
\subsection{Implications of the Number of Spiders}

One major  surprise of these \fermi\ searches has been the number of compact ($P_b < 1$~day) binary systems which show evidence of the pulsar wind interacting with the companion. Such compact, often eclipsing systems come in two varieties, now commonly referred to as `black widows' \citep[see][for a census]{Swihart22} and `redbacks' \citep[see][]{Strader19}.

In black widow systems, the pulsar's intense, relativistic wind is ablates the companion star, stripping it of mass, and potentially `devouring' it entirely over the course of billions of years\footnote{Hence the association with black widow spiders, who, according to some reports, sometimes eat their partner as part of the mating process.}.   PSR \bwa, with a degenerate $\sim$0.02\,\msun\ companion, was the first such system to be found and is considered the prototype of the class.  At radio wavelengths, this pulsar is eclipsed for $\sim$10\% of its orbit, likely due to screening by material blown off of the companion star (i.e., not occultation from the companion star itself). It also shows timing irregularities, especially around eclipse ingress/egress \citep{Fruchter1988}, commonly referred to as ``eclipse delays", as well as an orbital period derivative due to tidal effects.  Whether such systems can fully ablate their companion and turn into the observed isolated MSPs is a matter of ongoing debate.  Plausibly, some systems can ablate their companion star entirely, while others may not have sufficiently strong winds to do this within a Hubble time.

%Redbacks intro
Additionally, some eclipsing systems have companions which are significantly more massive ($\sim 0.2-0.4$\,\msun) and probably non-degenerate. These systems were first identified in globular clusters \citep[PSR~J1740$-$5340 in NGC6397;][]{dpa+01} and have now been termed redbacks\footnote{Redbacks are Australian spiders whose similar markings and habit of devouring mates make them cousins of the North American black widow.} from black widows \citep{rob11}.  Such systems possibly represent an earlier stage in the recycling process, as argued in the case of PSR~\spia, which has displayed the characteristics of both an accreting low-mass X-ray binary and a normal radio millisecond pulsar in the last decade \citep{asr+09,sah+14}. It is also possible that such systems have different progenitors and a different evolution than the black widows \citep[e.g.,][]{cct+13, bdh14, stf+15}.  Prior to \fermi, only four of the known MSPs in the Galactic field were in compact/eclipsing systems.
 
%---------------------------------------------------------------------------------
\subsubsection{The New Black Widows}

%Detection, properties of J0023
\textit{PSR~\mspb} has a spin period of 3.05\,ms and DM of 14.33\dm, with an implied DM-distance of 1.2\,kpc using the YMW16 model of \citet{ymw17}.~With $S_{350}$ $\sim$2\,mJy, the source was easily detected in a search of the first 200\,s of the observation. The orbital period is 3.3\,hr, with a minimum companion mass of 0.016\,\msun. These orbital properties are very similar to those of the black widow PSR~\bwc.  However, it does not show any evidence of an eclipse at either 350~MHz or 2~GHz. The radio profile is quite narrow ($W_{50} \sim 0.05$\,ms, see Figure~\ref{fig:profiles}), and despite its black-widow-like very low mass companion, extensive timing shows no evidence for orbital period variations.  

Early optical observations by \citet{breton2013} showed that the companion is strongly heated but that it appears to have a small radius that is significantly underfilling its Roche lobe. More extensive observations by \citet{mkc+23} show the Roche lobe fill factor is $\sim 50\%$. Therefore the surface gas may be too strongly gravitationally bound for the pulsar wind to ablate the surface, explaining the lack of eclipses. This may also explain why it does not exhibit orbital variations. The optical emission suggests a moderate inclination of $\sim 42^{\circ}$, and so even if the pulsar is as heavy as $2~M_{\odot}$, the companion would still be a very light $0.025~M_{\odot}$, and is therefore consistent with other black-widow masses. 

The narrow pulse profile and lack of orbital variations allow for high-precision timing and has led to this source being included in the NANOGrav and EPTA pulsar timing arrays \citep[e.g.,][]{vanHaasteren2011}.

%Detection, properties of J1124
\textit{PSR~\mspd} has a spin period of 2.41\,ms and DM of 44.86\dm, with an implied DM-distance of 1.0\,kpc. With $S_{350} \sim 0.3$\,mJy, it is one of the weaker sources presented here and required a search of the full 32-min discovery observation in order to be found.  The orbital period is 5.45\,hr, with a minimum companion mass of 0.028\,\msun.  
It is eclipsed around 40\% of the time at 350MHz, 20\% of the time at 820MHz, but apparently not eclipsed at 1400MHz. However, the 1400MHz TOAs  seem to have excess delays around superior conjunction.
As noted above, \citet{grm+14} found significant hard X-ray emission suggestive of an intrabinary shock. Optical observations of the system show that the companion is very strongly irradiated by high energy emission from the pulsar and/or shock \citep{drs+19}.

%Detection, properties of J1810
\textit{PSR~\mspa}  has a very short\footnote{It is the 4th fastest-spinning pulsar known in the Galactic field.} spin period of 1.66\,ms and DM of 39.66\dm, which implies a DM-distance of 2.4\,kpc. Relative to other known MSPs, the source is quite strong at low frequencies, $S_{350} = 20$\,mJy, and it was easily detected in a search of the first 200\,s of the observation. At even lower frequencies, it is one of the brightest pulsars in the sky \citep{kvh+16}. The orbital period is 3.6\,hr and the minimum companion mass is 0.045\,\msun.  Like the original black widow \bwa, \mspa\ has a very fast spin rate, though its orbital period is significantly shorter.  From 350 to 1500\,MHz, the radio profile evolves to have a sharp preceding peak (Figure~\ref{fig:profiles}), which is reminiscent of the trailing notch seen in the pulse profile of PSR B1937+21 \citep{McKee2019}. The $\gamma$-ray and radio peaks are roughly aligned (Figure \ref{fig:gamrad_j1810}), with a radio precursor becoming clearly detectable at 1500 MHz. 

\mspa\ is eclipsed for roughly 10\% of its orbital period, which is a similar eclipse fraction to the original black-widow \bwa\ \citep{Fruchter1988}. The eclipse boundaries are very sharp; the observed pulse intensity drops from its full non-eclipse value to undetectable within $\sim$2\,min. Eclipse delays, i.e. pulsed emission which has been shifted to later pulse longitude, are also observed at eclipse egress. Eclipse studies at lower frequencies with LOFAR and WEPSR show the eclipse edges to be highly variable, and that the mass loss rate is high enough to evaporate the companion within roughly a Hubble time \citep{pbc+18} ~\mspa\ has one of the highest minimum companion masses ($M_c>0.045$\,\msun) of all the known black widows. Extensive obervations with Keck indicate a very high neutron star mass of $M_{ns}\sim 2.1 M_{\odot}$ and a nearly Roche lobe filling companion with a mass $M_c\sim 0.065M_{\odot}$ \citep{rkf+21}. This  suggests ~\mspa\  may  be an evolutionary bridge between redbacks and black widows.

%---------------------------------------------------------------------------------
\subsubsection{The New Redbacks}

The longer eclipse fractions (typically 50\%) and higher companion masses ($0.2-0.4$\,\msun) of these systems identify them as redbacks, a class of system that before the discovery of the X-ray binary / MSP ``missing link" J1023$+$0038 \citep{asr+09} was only observed in globular clusters. The companions of these systems also show strong orbital modulation in optical \citep{breton2013} and X-ray \citep{grm+14, asr+09, Bogdanov15}. It is unclear whether the redbacks are an evolutionary precursor of the black widows.  

%Detection, properties of J2129
\textit{PSR~\mspe} has a spin period of 7.61\,ms and DM of 16.88\dm, with an implied DM-distance of 1.4\,kpc. With $S_{350} \sim 0.5$\,mJy, it also required the full 32-min observation to be detected.  It's high spin-down rate, which implies a larger dipole surface magnetic field $B\sim 1.6\times 10^9$ and a lower characteristic age than other redbacks. PSR~\mspe\ also has a relatively long orbital period ($P_B=15.2$~hr) and large minimum companion mass ($P_C \ga 0.37 M_{\odot}$). The high companion mass and relatively slow rotation period argue that this is a redback that is likely in an early stage of recycling.

The pulsar has large orbital period variations and is eclipsed $\sim 50\%$ of the time at low frequencies. The double peaked orbital X-ray light curve shows there is a strong shock which is apparently wrapped around the pulsar, which may indicate the companion magnetic field is dynamically important \citep{alnoori18}.  Optical observations show the companion is nearly Roche lobe filled, and also show long term trends in the brightness and temperature of the companion, suggesting the Roche lobe filling factor may be significantly changing on monthly to yearly time-scales \citep{bkb+16,alnoori18}. The system has a high inclination ($i\sim 77^{\circ}$), which is shown through both optical modeling and eclipses of the pulsar seen in $\gamma$-ray and X-ray data \citep{ckb+23}.

%Detection, properties of J2215
\textit{PSR~\mspc} has a spin period of 2.61\,ms and DM of 69.19\dm, with an implied DM-distance of 2.8\,kpc. It has the highest spin-down luminosity of any of the sources presented here and is the closest source to the Galactic plane.  The orbital period is 4.1\,hr, with a minimum companion mass of 0.21\,\msun. The pulsar has highly frequency dependant eclipses, with it being eclipsed for $\sim 1/2$ the orbit at 350~MHz but only for $\sim 1/4$ the orbit at 2~GHz. LOFAR studies have shown the eclipse fraction to increase to $\sim 3/4$ at 54MHz \citep{bfb+16}.  The orbital and eclipse properties are very similar to those of PSR~\spia, and consequently PSR~\mspc\ fits well within the redback category.  The radio pulse profile morphology shows strong evolution with frequency (Figure~\ref{fig:profiles}), which makes interpretation of the radio/$\gamma$-ray alignment challenging.

The \xmm{} light curves (Figures 6 and 7) show the double peaked orbital variations centered at pulsar inferior conjunction that seem to be a common feature of redbacks.   The non-thermal emission also has the very hard (photon index $\Gamma\sim 1$) typical of these sources. 

This source has garnered significant interest since optical studies have suggested a neutron star mass $M_{ns}>2.1 M_{\odot}$ \citep{lsc18,kr20} orbiting the Roche lobe filling $M_c \sim 0.33 M_{\odot}$ companion.
  
\textit{PSR~J1302--3205}, with an orbit of 18.8~hr,  minimum companion mass of $0.15M_{\odot}$, and evidence for a brief radio eclipse around superior conjunction, is potentially of the redback class of systems where the companion has not yet fully evolved and there is the potential for further accretion episodes. We obtained multiple images of PSR~J1302$-$3205 with the Las Cumbres Observatory (LCOGT) network of 1~m and 2~m telescopes using various filters. Using the XB-News pipeline analysis \citep{rbl+19}, there were no firm detections of the source. An observation with the Spectral Camera on the LCOGT 2~m  at Siding Springs taken near pulsar inferior conjunction, when heating should produce a maximum, puts a conservative upper limit magnitude of $m_{i^{\prime}} > 21.5$. An observation with the MuSCAT3 instrument on the LCOGT 2~m  at Haleakala  taken near pulsar superior conjunction put upper limits on the night side of the companion of $m_{i^{\prime}} > 22.3$, $m_{g^{\prime}}>22.3$, $m_{r^{\prime}}> 22.6$, and $m_{z_s}> 21.9$.   With an extinction estimate of $A_v\sim 0.22$ from \citet{dcl03} for the nominal distance of 1~kpc, the companion must be no brighter than a  M8 sub-dwarf. The minimum companion mass is consistent with that of a fully evolved helium white dwarf expected for the orbital period--companion mass relationship derived by several authors \citep[e.g.][]{ts99}. We further note that PSR~J1816$+$4510 has redback-like eclipses, but optical spectroscopy indicates the companion is more consistent with a proto-white dwarf \citep{Kaplan13}. Combined with modest, apparently thermal X-ray emission seen from the system, we find no convincing evidence that PSR~J1302$-$3205 should be classified as a redback at this time.  

%---------------------------------------------------------------------------------
\subsection{Implications for $\gamma$-ray Pulsar Population Statistics}

Our survey contains 22 MSPs (all known as of publication) which are apparently associated with \fermi-LAT sources.  Six are black widows, with the four discoveries reported here joined by PSR J2256$-$1024, discovered in the GBT drift scan survey \citep{csm+15}, and PSR J1805+0615, discovered in a survey of \fermi-LAT sources with Arecibo \citep{cck+16}.  Four are redbacks; and three are redback candidates \citep{Strader19}.  Four of these show evidence of intrabinary shocks in X-rays \citep{grm+14,rmg+15}, and five of them have been shown to significantly heat their companion \citep{breton2013, bellm2013}.  This might suggest that the remarkably high fraction of these systems in this $\gamma$-ray selected sample could be due to excess $\gamma$-ray emission from the pulsar wind interacting with the companion. However, it is clear from the estimated background levels in the $\gamma$-ray pulse profiles of our sample that essentially all of the observed $\gamma$-ray flux can be attributed to pulsed emission, and the $\gamma$-ray efficiencies (see Tables \ref{tab:timing_msps1} and \ref{tab:timing_msps2}) are compatible with those of other MSPs detected by \fermi-LAT (see 2PC and 3PC).

\begin{figure}
\begin{centering}
\includegraphics[scale=0.75,angle=0,width=0.98\linewidth]{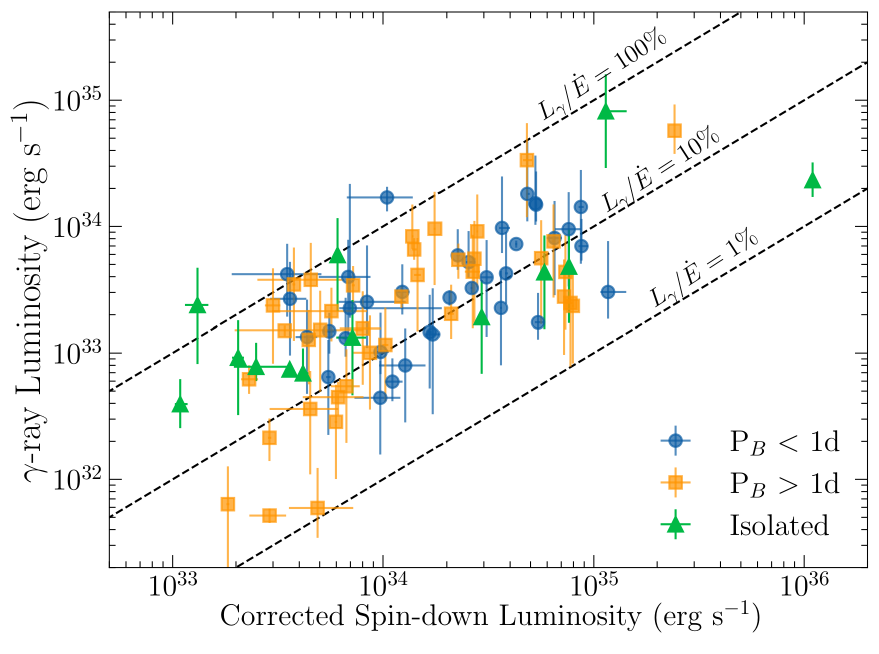} 
\caption{\label{fig:L-Edot}$\gamma$-ray luminosity ($L_{\gamma}$) as a function of 
corrected spin-down luminosity (\edot{}) for the sample of $\gamma$-ray MSPs reported in 3PC.}
\end{centering}
\end{figure}

We note that all four of our detected pulsars with $\dot E > 3\times 10^{34}$~erg s$^{-1}$ are eclipsing sources, suggesting that such sources are more energetic than the general MSP population. In general, $\gamma$-ray luminosity is correlated with \edot{}, albeit mildly for MSPs in the 10$^{33}$--10$^{34}$\,erg\,s$^{-1}$ range (see Figure \ref{fig:L-Edot} and 3PC).  This means that high-\edot{} MSPs will be detectable in $\gamma$ rays over a larger volume compared to low-\edot{} pulsars.  Conversely, radio luminosity is only weakly coupled with \edot{}. Moreover, the eclipsing systems are also harder to detect in radio pulsar surveys since their short orbital periods can require acceleration searches for even short observations, and they can be eclipsed for a significant fraction of the time.  Given these considerations, we expect $\gamma$-ray searches to be more efficient at detecting `spider' systems than radio surveys.

To examine the $\dot{E}$ dependence more concretely, we compared the measurements in 3PC, which provide the best-available estimates of $\gamma$-ray luminosity and Shklovskii-corrected $\dot{E}$, for isolated, short-period, and long-period binaries, with results shown in Figure \ref{fig:L-Edot}.  Visually, there seems to be a lack of low-$\dot{E}$ short-period binaries.  However, a two-sample Kolmogorov-Smirnov test indicates only modest evidence for a difference between the binary and isolated populations (p-value 5\%) and no evidence for a difference between short- and long-period binaries (p-value 15\%).

Of the 19 sources that we observed that are listed in the 3FGL catalog and remain unidentified or unassociated with possible AGN counterparts, most remain viable pulsar candidates. If they are MSPs, their non-detection is likely due to large accelerations due to orbital motion, large eclipse fractions, or simply radio faintness, as most MSPs appear to have radio and $\gamma$-ray beams sweeping similar regions of the sky, with $> 75$\% of $\gamma$-ray MSPs being radio-loud (3PC), although in the case of black widows and redbacks they may be eclipsed more than half of the time at low frequencies. Redbacks may also spend some fraction of their time in a low accretion state with no observable radio pulsations but can have observable unpulsed $\gamma$-ray emission in this state \citep{sah+14, jrr+15}. We therefore estimate that about 5 of the unidentified sources are MSPs. Some of these may be detectable with re-observations or at higher frequencies (note that PSR~J1921+0137, discovered in an 820-MHz survey, is undetectable in our data). In fact, three of our targets, 1FGL~J0523.5$-$2529 \citep{scs+14}, 1FGL~J1119.9$-$2205 \citep{Swihart22b},  and 1FGL\,J1544.5$-$1127 \citep{Strader19}, have been suggested to be redbacks through the identification of a plausible optical counterpart.  PSR~J0955$-$3947, another redback candidate \citep{Braglia20}, was recently confirmed as a short-orbit pulsar by TRAPUM\footnote{http://trapum.org/discoveries/} \citep{cbb+23}.

Three of our targets are now known to be non-recycled, middle-aged pulsars discovered through blind $\gamma$-ray searches.  These appear to be radio-quiet pulsars, and none are detectable in our radio data.  Compared to the 15 sources that have not yet been identified in our sample, these three sources are on average 2.6 times brighter in $\gamma$ rays. Assuming the $\gamma$-ray blind search detectability is largely flux limited and that there
is a roughly spherical distribution of nearby, high Galactic latitude pulsars, this would suggest that roughly a dozen of these fainter sources could be middle-aged ($\tau$$\sim$$10^5$~yr), radio-quiet pulsars. We note that nearby $\gamma$-ray pulsars are likely to have fairly low spin-down powers and hence are much more likely to be radio faint, compared to younger pulsars deep in the plane (3PC). We therefore find it plausible that essentially all of our remaining unassociated sources are either MSPs or nearby middle-aged pulsars.

\citet{jvh+14} divide $\gamma$-ray pulsars into three classes based on their pulse profiles: $\gamma$-ray peaks trail radio (Class I); are aligned (Class II); $\gamma$-ray peaks leading radio (Class III).  (See also \citet{2013MNRAS.430..571E} for a similar classification scheme.)  Outer gap and slot gap (two-pole caustic) models best fit roughly equal numbers of Class I and III, while Class II are exclusively fit with pair-starved polar cap models. According to this scheme,  PSR~J0102+4839 is a Class III pulsar, with the $\gamma$-ray peak leading the radio peak. PSR~J0307+7443 is a Class II, with the $\gamma$-ray and radio peaks aligned. PSR~J0340+4130 is a Class III, with the primary $\gamma$-ray peak leading the primary radio peak, though it should be noted that the primary radio peak is aligned with the secondary $\gamma$-ray peak. \citet{jvh+14} found that Class II MSPs had the shortest spin periods, with five of the six members of this class having spin periods less than 2 ms, while Class I and Class III MSPs had longer spin periods. PSR~J0307+7443, with a spin period of 3.2 ms, does not fit that trend.

The close-orbit pulsars are more difficult to classify since nearly all show strong frequency dependence in their radio pulse profiles.  
At 350~MHz, PSR~J1302$-$3258 is a Class I, with the main radio peak leading the $\gamma$-ray peak.~However, we note that an interpulse becomes apparent at 820 MHz which trails the secondary $\gamma$-ray peak and becomes even more prominent at higher radio frequencies. The \fermi-LAT 2PC catalog used the brightest profile peak (at 2~GHz) of PSRs~J0023+0923, J1810+1744, and J2215+5135 when calculating phase lags. However, both PSRs~J0023+0923 and J1810+1744  show a narrow ``pre-cursor'' peak that is not visible at 350~MHz but whose peak brightness begins to approach the main peak's brightness by 2 GHz, and would presumably become the ``main peak'' at higher frequencies. At low frequencies, PSR~J0023+0923 is a Class I and  PSR~J1810+1744 a Class II, but if measured from the flatter spectrum pre-cursor peak, the classifications would be reversed.~The pulse profile evolution of PSR~J2215+5135 is even more pronounced. At 350~MHz, it has two peaks which are fairly well aligned with the $\gamma$-ray peaks, implying this would be a Class II source, but at 2~GHz there is a single peak that is neither of the 350-MHz peaks and which trails the $\gamma$-ray peaks, and so would be a Class III. Pulse profiles which strongly evolve with frequency seem to be a feature of many black widows and redbacks. We note that this has also been noted in the original black widow pulsar B1957+20 \citep{jvh+14} as well as in the redbacks PSR J1048+2339 \citep{drc+16} and PSR J1023+0038 \citep{asr+09}. For these systems, at least, classification schemes based on relative phases of $\gamma$-ray and radio peaks could break down. The pulse-profile evolution common to these also brings up the possibility that the companion is somehow influencing the radio emission of the pulsar if this frequency dependence is truly more common among these systems than in the general millisecond pulsar population. 

%---------------------------------------------------------------------------------
\subsection{Implications for General Millisecond Pulsar Population Statistics}

In the standard theory of MSP formation, a neutron star in a binary system is `recycled' by accreting matter from its companion star \citep{acr+82,sta05}, and consequently, most MSPs are expected to be in binary systems with a white dwarf, the natural end state of the evolution of the light companion following accretion and spinup.  However, \textit{psrcat} currently\footnote{MSP class counts obtained with \textit{psrcat} use version 1.70 and require period $P<10$\,ms and a well-measured $\dot{P}$.} lists 52 MSPs 
in the Galactic field without binary companions, about 23\% of the total. The formation of these isolated MSPs is still not well understood, though one possibility is that they ablated their companions with their strong relativistic particle winds \citep{rst89}. Another $\sim$32 MSPs\footnote{\url{https://apatruno.wordpress.com/about/millisecond-pulsar-catalogue/}} in the Galactic field do not have ordinary white dwarf companions but instead have either under-massive companions or non-degenerate companions. Understanding the various sub-populations of MSPs is critical for improving our view of the pulsar recycling process and whether it proceeds via only one channel or a variety of routes.  The rapid rate of MSP discovery is can greatly help in this endeavor by rounding out the population with respect to previous observational biases.

Deep searches targeting $\gamma$-ray sources have a very different set of observational biases than shallow, wide-field radio surveys. Due to there being a much stronger correlation between spin-down luminosity and apparent $\gamma$-ray luminosity than there is with apparent radio luminosity, a $\gamma$-ray targeted survey will be more biased towards nearby, energetic MSPs and less biased towards high radio luminosities. Older low-frequency surveys had much less spectral resolution and were, therefore, more susceptible to DM smearing of the pulse profile, degrading their sensitivity to MSPs. Higher DM and scattering in the Galactic plane would also inhibit the discovery of MSPs. In the past, there were also strong biases against detecting pulsars with short orbital periods since acceleration searches only became common practice after ~2000.

Only two out of 18 MSPs (11\%) detected in this survey are isolated.  3PC reports 91 MSPs discovered by targeting unidentified Fermi sources, and of these, only 16 are isolated\footnote{We exclude two MSPs which were discovered in a targeted search of LAT data \citep{clark18}.}, about 18\%.  Both are lower than the roughly 25\% isolated fraction of the Galactic plane population found before 2009. Furthermore, eight of the pulsars detected in this survey have orbital periods $P_b < 1$\,d.  In the same 3PC sample, the fraction of short-period binaries is 54\%.  In stark contrast, only $\sim$14\% of MSPs found before 2009 have such short binary periods.  As a result of the \fermi\ searches, there are now more known MSPs with short orbits than isolated ones. If we compare the Galactic field population of $P_b < 1$~day MSPs to the ones in globular clusters, in both cases, the fraction is now $\sim$25\%. This calls into question the importance of binary exchange in the formation of these systems in globular clusters. 
 
\begin{figure}
\centering
\epsscale{.80}
\includegraphics[scale=0.75,width=0.98\linewidth]{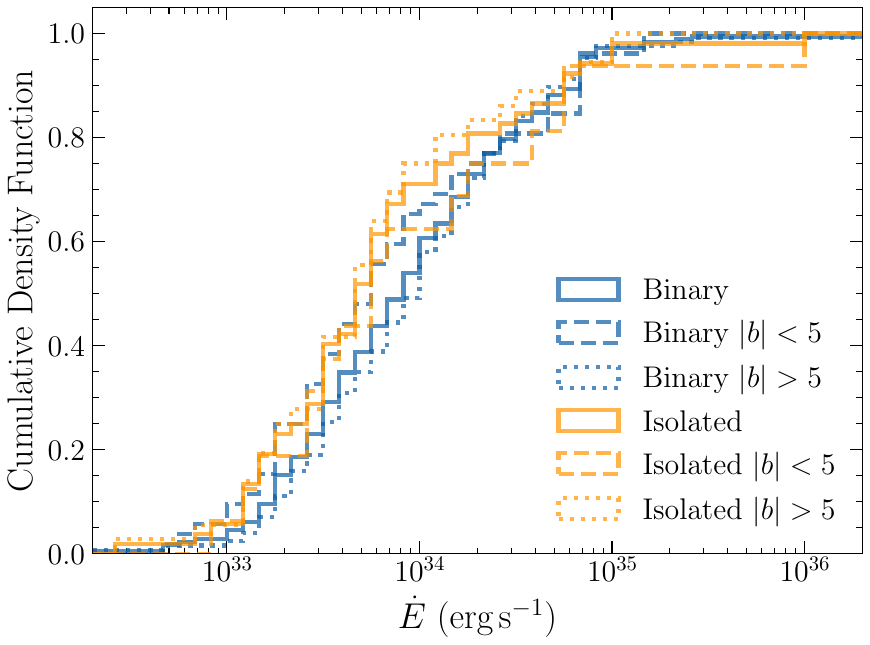}
\includegraphics[scale=0.75,width=0.98\linewidth]{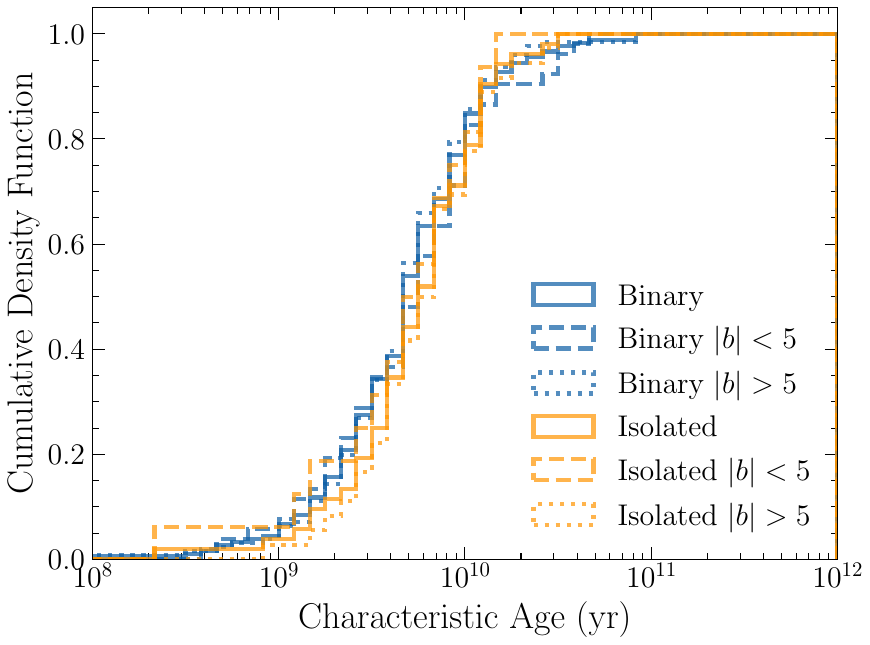}
\caption{The cumulative distribution functions (CDFs) of the spin-down energy loss rates (top) and ages (bottom) for isolated and binary MSPs. The different line styles show the full, low-latitutde (dashed) and high-latitude (dotted) samples.  These CDFs are broadly compatible, but a Kolmogorov-Smirnov test provides modest evidence that the high-latitude $\dot{E}$ distributions differ.  See main text for details of the sample construction and analysis.}
 \label{fig:CDFs}
\end{figure}

Compared to binary MSPs, isolated MSPs could in principle be older or sustain a higher spin-down energy loss ($\dot{E}$) in order to fully disintegrate their companions.  To investigate this, we have compared the $\dot{E}$ and characteristic ages of binary and isolated MSPs both in full and restricted to Galactic latitudes $|b| > 5^{\circ}$.  In order to obtain a larger sample, we use values of $\dot{P}$ which have not been corrected for the Shklovskii effect to estimate $\dot{E}$ and the characteristic age. (In addition to this caveat, it is important to note that these are derived quantities of $P$ and $\dot{P}$ and are only rough proxies of the true ages and spindown luminosities.) The distribution function for these two quantities is shown in Figure \ref{fig:CDFs}, inspection of which indicates that the distributions for isolated and binary MSPs is broadly compatible.  We quantitatively compared the distributions using a two-sample Kolmogorov-Smirnov test, finding modest evidence (p-value 5\%) that the $\dot{E}$ distributions differ.  Further examination indicates that the $|b|>5$ pulsars have only a 2\% probability of being drawn from the same distribution, with isolated pulsars generally having lower values of $\dot{E}$, while the $|b|<5$ populations are compatible.  There is no substantial difference in any of the age distributions.  Further work will be necessary to determine if the observed difference in spindown luminosities is intrinsic or an observational selection effect against low-$\dot{E}$, high-latitude binaries.
    
%---------------------------------------------------------------------------------
\section{Conclusion}

This work reports on a groundbreaking survey conducted more than ten years ago.  Although we wish it had been published sooner, much of the delay was due to the success of the project.  The MSP discoveries necessitated timing observations, and the discovery of such an unusually---at the time---large number of `spider' pulsars demanded multiwavelength follow-up.  It was always tempting to acquire more observational data and better understand this growing source class and its relation to $\gamma$ rays.  The haul of new MSPs also created its own distractions, because it inspired numerous follow-up surveys, many cited above, of different parts of the sky, at different frequency ranges, and with different \fermi{} source lists.  The success of this method continues through the present day, with powerful new instruments like FAST and MeerKAT and finding ever fainter radio pulsars in \fermi{} sources \citep{wlc+21,cbb+23}.

Despite the long gap between observations and report, this work is the definitive record of the radio observations carried out; the follow-up timing observations; and the phase-aligned $\gamma$-ray and radio pulse profiles.  Together with a synthesis of multi-wavelength observations, particularly X-ray but also benefiting from optical analyses of redback candidates that lack pulsations, we can argue that $\gamma$-ray targeted searches correct a long-standing bias in radio surveys, which seem to have overproduced isolated MSPs relative to the general population. And it is only with the benefit of the substantial followup that we can appreciate how efficient the original target selection was.  It is thus fitting that this report finally appear in the literature, and we conclude: better late than never!

%---------------------------------------------------------------------------------
\begin{acknowledgments}

This work was partially supported by NASA Fermi Guest Investigator Program grants NNG10PB13P, NNX10AP81G, NNX12AE32G, NNX12AP21G, and NNX13AO96G. The Green Bank Telescope is operated by the National Radio Astronomy Observatory, a facility of the National Science Foundation operated under cooperative agreement by Associated Universities, Inc. The WSRT is operated by ASTRON/NWO. This work makes use of observations from the Las Cumbres Observatory global telescope network, with support from New York University Abu Dhabi. This paper includes observations made with the MuSCAT3 instrument, developed by Astrobiology Center and under financial supports by JSPS KAKENHI (JP18H05439) and JST PRESTO (JPMJPR1775), at Faulkes Telescope North on Maui, HI, operated by the Las Cumbres Observatory. We thank Gyula Jozsa for help scheduling the WSRT observations. JWTH acknowledges funding from an NWO Vidi fellowship and ERC Starting Grant ``DRAGNET'' (337062). MSER acknowledges funding from the NYU Abu Dhabi faculty research fund. Work at NRL is supported by NASA and by ONR 6.1 basic research funding. ECF is supported by NASA under award number 80GSFC21M0002. SMR is a CIFAR Fellow. MAM is supported by NSF award 2009425. MED, ECF, MK, MAM, SMR, and PSR and members of the NANOGrav Physics Frontiers Center supported by NSF award 2020265.

The \textit{Fermi} LAT Collaboration acknowledges generous ongoing support
from a number of agencies and institutes that have supported both the
development and the operation of the LAT as well as scientific data analysis. These include the National Aeronautics and Space Administration and the Department of Energy in the United States, the Commissariat \`a l'Energie Atomique and the Centre National de la Recherche Scientifique / Institut National de Physique Nucl\'eaire et de Physique des Particules in France, the Agenzia Spaziale Italiana and the Istituto Nazionale di Fisica Nucleare in Italy, the Ministry of Education, Culture, Sports, Science and Technology (MEXT), High Energy Accelerator Research
Organization (KEK) and Japan Aerospace Exploration Agency (JAXA) in Japan, and the K.~A.~Wallenberg Foundation, the Swedish Research Council and the Swedish National Space Board in Sweden.

Partially based on observations obtained with \textit{XMM-Newton}, an ESA science mission with instruments and contributions directly funded by ESA Member States and NASA
 
Additional support for science analysis during the operations phase is gratefully acknowledged from the Istituto Nazionale di Astrofisica in Italy and the Centre National d'\'Etudes Spatiales in France. This work performed in part under DOE Contract DE-AC02-76SF00515.

We thank the NANOGrav collaboration for contributing data on PSR~J0340+4130 that were used for this paper. We also thank Duncan Lorimer and Elizabeth Ferrara for maintaining the list of published and unpublished Galactic field MSPs.
\end{acknowledgments}

%---------------------------------------------------------------------------------
\facilities{
Green Bank Telescope (GBT),
Fermi Gamma-ray Space Telescope (Fermi),
Giant Metrewave Radio Telescope (GMRT), 
Westerbork Synthesis Radio Telescope (WSRT),
Nan\c{c}ay Radio Telescope (NRT),
Neil Gehrels Swift Observatory (Swift),
Chandra X-ray Observatory (CXO),
X-ray Multi-Mirror Mission (XMM-Newton),
}

%---------------------------------------------------------------------------------
% References

%---------------------------------------------------------------------------------

\begin{thebibliography}{}

\bibitem[Abdo et al.(2013)]{2PC} Abdo, A.~A., Ajello, M., and Allafort, A., et al.\ 2013, \apjs, 208, 17
\bibitem[Smith et al.(2023)]{3PC} Smith, D.~A., Abdollahi, S., and Ajello, M., et al.\ 2023, \apj, 958, 2

\bibitem[Abdo et al.(2010a)]{aaa+10a} Abdo, A.~A., Ackermann, 
M., Ajello, M., et al.\ 2010, \apjs, 187, 460

\bibitem[Abdo et al.(2010b)]{aaa+10b} Abdo, A.~A., Ackermann, 
M., Ajello, M., et al.\ 2010, \apjs, 188, 405 

\bibitem[Abdo et al.(2010c)]{aaa+10c} Abdo, A.~A., Ackermann, 
M., Ajello, M., et al.\ 2010, \apj, 715, 429

\bibitem[Abdo et al.(2009a)]{aaa+09a} Abdo, A.~A., Ackermann, 
M., Ajello, M., et al.\ 2009, Science, 325, 840

\bibitem[Abdo et al.(2009b)]{aaa+09b} Abdo, A.~A., Ackermann, 
M., Ajello, M., et al.\ 2009, Science, 325, 848

\bibitem[Abdo et al.(2009c)]{aaa+09c} Abdo, A.~A., Ackermann, 
M., Ajello, M., et al.\ 2009, \apj, 700, 597 

\bibitem[Abdo et al.(2013)]{abdo13} Abdo, A.~A., Ajello, M., Allafort, A., et al.\ 2013, \apjs, 208, 17. doi:10.1088/0067-0049/208/2/17

\bibitem[Abdollahi et al.(2020)]{4fgl} Abdollahi, S., Acero, F., Ackermann, M., et al.\ 2020, \apjs, 247, 33. doi:10.3847/1538-4365/ab6bcb
 
\bibitem[Ackermann et al.(2012)]{P7Paper} Ackermann, M., Ajello, M., Albert, A., et al.\ 2012, \apjs, 203, 4

\bibitem[FERMI-LAT Collaboration et al.(2022)]{2022Sci...376..521F} Ajello, M., Atwood, W.~B., et al.\ 2022, Science, 376, 521. doi:10.1126/science.abm3231

\bibitem[Allafort et al.(2013)]{allafort13} Allafort, A., Baldini, L., Ballet, J., et al.\ 2013, \apjl, 777, L2. doi:10.1088/2041-8205/777/1/L2

\bibitem[Al Noori et al.(2018)]{alnoori18} Al Noori, H., Roberts, M.~S.~E., Torres, R.~A., et al.\ 2018, \apj, 861, 89. doi:10.3847/1538-4357/aac828

\bibitem[Alpar et al.(1982)]{acr+82} Alpar, M.~A., Cheng, 
A.~F., Ruderman, M.~A., \& Shaham, J.\ 1982, \nat, 300, 728 

\bibitem[Archibald et al.(2009)]{asr+09} Archibald, A.~M., Stairs, I.~H., Ransom, S.~M., et al.\ 2009, Science, 324, 1411 

\bibitem[Arzoumanian et al.(2015)]{abb+15} Arzoumanian, Z., 
Brazier, A., Burke-Spolaor, S., et al.\ 2015, arXiv:1505.07540

\bibitem[Atwood et al.(2009)]{Atwood09} Atwood, W.~B., Abdo, A.~A., 
Althouse, W., et al.\ 2009, \apj, 697, 1071

\bibitem[{Backer {et~al.}(1982)Backer, Kulkarni, Heiles, Davis, \&
  Goss}]{bkh+82}
Backer, D.~C., Kulkarni, S.~R., Heiles, C., Davis, M.~M., \& Goss, W.~M. 1982,
  Nature, 300, 615
  
\bibitem[Barr et al.(2013)]{bgc+13} Barr, E.~D., Guillemot, 
L., Champion, D.~J., et al.\ 2013, \mnras, 429, 1633

\bibitem[Bellm et al.(2013)]{bellm2013} Bellm, E., Djorgovski, 
S.~G., Drake, A.~J., et al.\ 2013, American Astronomical Society Meeting 
Abstracts \#221, 221, \#154.10 

\bibitem[Bellm et al.(2016)]{bkb+16} Bellm, E.~C., Kaplan, D.~L., Breton, R.~P., et al.\ 2016, \apj, 816, 74. doi:10.3847/0004-637X/816/2/74


\bibitem[Benvenuto et al.(2014)]{bdh14} Benvenuto, O.~G., De Vito, M.~A., \& Horvath, J.~E.\ 2014, \apjl, 786, L7 

\bibitem[Bhattacharyya et al.~(2013)]{bhat2013}~Bhattacharyya, B., Roy, J.,  Gupta, Y., et al.\ 2013, \apjl, 773, L12

\bibitem[Bhattacharyya et al.~(2021)]{bhat21}~Bhattacharyya, B., Roy, J.,  Johnson, T.~J., et al.\ 2021, \apj, 910, 2
 
\bibitem[Bogdanov et al.(2015)]{Bogdanov15} Bogdanov, S., Archibald, A.~M., Bassa, C., et al.\ 2015, \apj, 806, 148

\bibitem[Bognar et al.(2015)]{bognar} Bognar, K., Roberts, M., 
\& Chatterjee, S.\ 2015, American Astronomical Society Meeting Abstracts, 225, \#346.11 

\bibitem[Braglia et al.(2020)]{Braglia20} Braglia, C., Mignani, R.~P., Belfiore, A., et al.\ 2020, \mnras, 497, 5364. doi:10.1093/mnras/staa2339

\bibitem[Breton et al.(2013)]{breton2013} Breton, R.~P., van 
Kerkwijk, M.~H., Roberts, M.~S.~E., et al.\ 2013, \apj, 769, 108 

\bibitem[Broderick et al.(2016)]{bfb+16} Broderick, J.~W., Fender, R.~P., Breton, R.~P., et al.\ 2016, \mnras, 459, 2681. doi:10.1093/mnras/stw794

\bibitem[Burrows et al.(2000)]{bhn+00} Burrows, D.~N., Hill, 
J.~E., Nousek, J.~A., et al.\ 2000, \procspie, 4140, 64 

\bibitem[Camilo et al.(2012)]{ckr+12} Camilo, F., Kerr, M., 
Ray, P.~S., et al.\ 2012, \apj, 746, 39 

\bibitem[Camilo et al.(2015)]{ckr+15} Camilo, F., Kerr, M., Ray, P.~S., et al.\ 2015, \apj, 810, 85. doi:10.1088/0004-637X/810/2/85


\bibitem[Camilo et al.(2016)]{cam16} Camilo, F., Reynolds, J.~E., Ransom, S.~M., et al.\ 2016, \apj, 820, 6. doi:10.3847/0004-637X/820/1/6

\bibitem[Champion et al.(2005)]{cml05} Champion, D.~J., 
McLaughlin, M.~A., \& Lorimer, D.~R.\ 2005, \mnras, 364, 1011

\bibitem[Chen et al.(2013)]{cct+13} Chen, H.-L., Chen, X., Tauris, T.~M., \& Han, Z.\ 2013, \apj, 775, 27 

\bibitem[Clark et al.(2018)]{clark18} Clark, C.~J., Pletsch, H.~J., Wu, J., et al.\ 2018, Science Advances, 4, eaao7228. doi:10.1126/sciadv.aao7228

\bibitem[Clark et al.(2020)]{clark20} Clark, C.~J., Nieder, L., Voisin, G., et al.\ 2020, \mnras. doi:10.1093/mnras/staa3484

\bibitem[Clark et al.(2023a)]{cbb+23} Clark, C.~J., Breton, R.~P., Barr, E.~D., et al.\ 2023, \mnras, 519, 5590. doi:10.1093/mnras/stac3742

\bibitem[Clark et al.(2023b)]{ckb+23} Clark, C.~J., Kerr, M., Barr, E.~D., et al.\ 2023, Nature Astronomy, 7, 451. doi:10.1038/s41550-022-01874-x

\bibitem[Cognard et al.(2011)]{cgj+11} Cognard, I., Guillemot, 
L., Johnson, T.~J., et al.\ 2011, \apj, 732, 47

\bibitem[Cordes 
\& Lazio(2002)]{cl02} Cordes, J.~M., \& Lazio, T.~J.~W.\ 2002, arXiv:astro-ph/0207156

\bibitem[Crawford et al.(2006)]{crh+06} Crawford, F., Roberts, 
M.~S.~E., Hessels, J.~W.~T., et al.\ 2006, \apj, 652, 1499

\bibitem[Cromartie et al.(2016)]{cck+16} Cromartie, H.~T., Camilo, F., Kerr, M., et al.\ 2016, \apj, 819, 34 

\bibitem[Crowter et al.(2015)]{csm+15} Crowter, K., Stairs, 
I.~H., McPhee, C.~A., et al.\ 2015, American Astronomical Society Meeting 
Abstracts, 225, \#307.01 

\bibitem[D'Amico et al.(2001)]{dpa+01} D'Amico, N., Possenti, A., Manchester, R.~N., et al.\ 2001, \apjl, 561, L89 

\bibitem[Deneva et al.(2016)]{drc+16} Deneva, J.~S., Ray, P.~S., Camilo, F., et al.\ 2016, \apj, 823, 105 

\bibitem[Deneva et al.(2021)]{den21} Deneva, J.~S., Ray, P.~S., Camilo, F., et al.\ 2021, \apj, 909, 6

\bibitem[Demorest et al.(2010)]{dpr+10} Demorest, P.~B., 
Pennucci, T., Ransom, S.~M., Roberts, M.~S.~E., 
\& Hessels, J.~W.~T.\ 2010, \nat, 467, 1081

\bibitem[Demorest et al.(2013)]{dfg+13} Demorest, P.~B., 
Ferdman, R.~D., Gonzalez, M.~E., et al.\ 2013, \apj, 762, 94 

\bibitem[Draghis et al.(2019)]{drs+19} Draghis, P., Romani, R.~W., Filippenko, A.~V., et al.\ 2019, \apj, 883, 108. doi:10.3847/1538-4357/ab378b

\bibitem[Drimmel et 
al.(2003)]{dcl03} Drimmel, R., Cabrera-Lavers, A., \& L{\'o}pez-Corredoira, M.\ 2003, \aap, 409, 205 

\bibitem[DuPlain et al.(2008)]{drd+08} DuPlain, R., Ransom, 
S., Demorest, P., et al.\ 2008, \procspie, 7019,

\bibitem[Edmonds et al.(2002)]{egc+02} Edmonds, P.~D., Gilliland, R.~L., Camilo, F., Heinke, C.~O., \& Grindlay, J.~E.\ 2002, \apj, 579, 741

\bibitem[Espinoza et al.(2013)]{2013MNRAS.430..571E} Espinoza, C.~M., Guillemot, L., {\c{C}}elik, {\"O}., et al.\ 2013, \mnras, 430, 571. doi:10.1093/mnras/sts657

\bibitem[Falcone et al.(2011)]{fsf+11} Falcone, A., Stroh, M., 
Ferrara, E., et al.\ 2011, AAS/High Energy Astrophysics Division, 12, 
\#04.03

\bibitem[Fruchter et al.(1988)]{Fruchter1988} Fruchter, A.~S., Stinebring, D.~R., Taylor, J.~H.\ 1990, Nature, 333, 237 


\bibitem[Fruchter et al.(1990)]{fbb+90} Fruchter, A.~S., Berman, G., Bower, G., et al.\ 1990, \apj, 351, 642 

\bibitem[Gentile et al.(2014)]{grm+14} Gentile, P.~A., 
Roberts, M.~S.~E., McLaughlin, M.~A., et al.\ 2014, \apj, 783, 69 

\bibitem[Green et al.(2015)]{gsf+15} Green, G.~M., Schlafly, E.~F., Finkbeiner, D.~P., et al.\ 2015, \apj, 810, 25. doi:10.1088/0004-637X/810/1/25

\bibitem[Gonthier et 
al.(2005)]{ggh+05} Gonthier, P.~L., Guilder, R., Harding, A.~K., Grenier, I.~A., \& Perrot, C.~A.\ 2005, \apss, 297, 71

\bibitem[Guillemot et al.(2012)]{gfc+12} Guillemot, L., 
Freire, P.~C.~C., Cognard, I., et al.\ 2012, \mnras, 422, 1294

\bibitem[G{\"u}ver \& {\"O}zel(2009)]
{go09} G{\"u}ver, T., \& {\"O}zel, F.\ 2009, \mnras, 400, 2050

\bibitem[Halpern et al.(2001)]{hcg+01} Halpern, J.~P., Camilo, 
F., Gotthelf, E.~V., et al.\ 2001, \apjl, 552, L125

\bibitem[Halpern et al.(2008)]{hcg+08} Halpern, J.~P., Camilo, 
F., Giuliani, A., et al.\ 2008, \apjl, 688, L33

\bibitem[Haslam et 
al.(1982)]{hss+82} Haslam, C.~G.~T., Salter, C.~J., Stoffel, H., \& Wilson, W.~E.\ 1982, \aaps, 47, 1 

\bibitem[Hessels et al.(2005)]{hrr+05} Hessels, J., Ransom, 
S., Roberts, M., et al.\ 2005, Binary Radio Pulsars, 328, 395

\bibitem[Hessels et al.(2011)]{hrm+11} Hessels, J.~W.~T., 
Roberts, M.~S.~E., McLaughlin, M.~A., et al.\ 2011, American Institute of 
Physics Conference Series, 1357, 40 

\bibitem[Hobbs et al.(2006)]{hem06} Hobbs, G.~B., Edwards, 
R.~T., \& Manchester, R.~N.\ 2006, \mnras, 369, 655

\bibitem[Hobbs et al.(2010)]{haa+10} Hobbs, G., Archibald, A., 
Arzoumanian, Z., et al.\ 2010, Classical and Quantum Gravity, 27, 084013 

\bibitem[Hotan et al.(2004)]{hvm04} Hotan, A.~W., van 
Straten, W., \& Manchester, R.~N.\ 2004, \pasa, 21, 302 

\bibitem[Hui et al.(2015)]{hhp+15} Hui, C.~Y., Hu, C.~P., 
Park, S.~M., et al.\ 2015, \apjl, 801, L27

\bibitem[Jenet et al.(2005)]{jcl05} Jenet, F.~A., Creighton, 
T., \& Lommen, A.\ 2005, \apjl, 627, L125 

\bibitem[Johnson et al.(2014)]{jvh+14} Johnson, T.~J., Venter, 
C., Harding, A.~K., et al.\ 2014, \apjs, 213, 6 

\bibitem[Johnson et al.(2015)]{jrr+15} Johnson, T.~J., Ray, 
P.~S., Roy, J., et al.\ 2015, \apj, 806, 91

\bibitem[Kandel \& Romani(2020)]{kr20} Kandel, D. \& Romani, R.~W.\ 2020, \apj, 892, 101. doi:10.3847/1538-4357/ab7b62

\bibitem[Kaplan et al.(2013)]{Kaplan13} Kaplan, D.~L., Bhalerao, V.~B., van Kerkwijk, M.~H., et al.\ 2013, \apj, 765, 158. doi:10.1088/0004-637X/765/2/158

\bibitem[Karuppusamy et al.(2008)]{Karuppusamy2008} Karuppusamy, R., Stappers, B., van Straten, W.\ 2008, PASP, 120, 191

\bibitem[Keith et al.(2008)]{kjk+08} Keith, M.~J., Johnston, 
S., Kramer, M., et al.\ 2008, \mnras, 389, 1881

\bibitem[Keith et al.(2011)]{kjr+11} Keith, M.~J., Johnston, 
S., Ray, P.~S., et al.\ 2011, \mnras, 414, 1292

\bibitem[Kerr(2011)]{Kerr11} Kerr, M.\ 2011, \apj, 732, 38

\bibitem[Kerr et al.(2012)]{kcj+12} Kerr, M., Camilo, F., 
Johnson, T.~J., et al.\ 2012, \apjl, 748, L2

\bibitem[Kerr et al.(2015)]{kerr15} Kerr, M., Ray, P.~S., Johnston, S., et al.\ 2015, \apj, 814, 128. doi:10.1088/0004-637X/814/2/128

\bibitem[Kondratiev et al.(2016)]{kvh+16} Kondratiev, V.~I., Verbiest, J.~P.~W., Hessels, J.~W.~T., et al.\ 2016, \aap, 585, A128. doi:10.1051/0004-6361/201527178

\bibitem[Kramer et al.(1998)]{kxl+98} Kramer, M., Xilouris, 
K.~M., Lorimer, D.~R., et al.\ 1998, \apj, 501, 270

\bibitem[Kuiper et al.(2000)]{khv+00} Kuiper, L., Hermsen, W., Verbunt, F., et al.\ 2000, \aap, 359, 615



\bibitem[Linares et al.(2018)]{lsc18} Linares, M., Shahbaz, T., \& Casares, J.\ 2018, \apj, 859, 54. doi:10.3847/1538-4357/aabde6

\bibitem[Loredo (1992)]{l92} Loredo, T.J.\ 1992, In: Feigelson, E.D., Babu, G.J. (eds) Statistical Challenges in Modern Astronomy. Springer, New York, NY.\ 275. doi:10.1007/978-1-4613-9290-3\_31

\bibitem[Lundgren et al.(1995)]{Lundgren95} Lundgren, S.~C., Zepka, A.~F., \& Cordes, J.~M.\ 1995, \apj, 453, 419. doi:10.1086/176402

\bibitem[Luo et al.(2021)]{luo21} \bibitem[Luo et al.(2021)]{2021ApJ...911...45L} Luo, J., Ransom, S., Demorest, P., et al.\ 2021, \apj, 911, 45. doi:10.3847/1538-4357/abe62f

\bibitem[Manchester et al.(2005)]{psrcat} Manchester, R.~N., Hobbs, G.~B., Teoh, A., et al.\ 2005, \aj, 129, 1993.

\bibitem[Marelli et al.(2011)]{Marelli11} Marelli, M., De Luca, A., \& Caraveo, P.~A.\ 2011, \apj, 733, 82. doi:10.1088/0004-637X/733/2/82


\bibitem[Mata S{\'a}nchez et al.(2023)]{mkc+23} Mata S{\'a}nchez, D., Kennedy, M.~R., Clark, C.~J., et al.\ 2023, \mnras, 520, 2217. doi:10.1093/mnras/stad203


\bibitem[Matthews et al.(2016)]{mnf+16} Matthews, A.~M., Nice, D.~J., Fonseca, E., et al.\ 2016, \apj, 818, 92 

\bibitem[Mayer et al.(2013)]{mbh+13} Mayer, M., Buehler, R., 
Hays, E., et al.\ 2013, \apjl, 775, L37

\bibitem[McKee et al.(2019)]{McKee2019} McKee, J.~W., Stappers, B.~W., Bassa, C.~G., et al.\ 2019, \mnras, 483, 4784  

\bibitem[McLaughlin et al.(1996)]{mmct96} McLaughlin, M.~A.,  
Mattox, J.~R., Cordes, J.~M., \& Thompson, D.~J.\ 1996, \apj, 473, 763  

\bibitem[The NANOGrav Collaboration et al.(2015)]{nano15} The NANOGrav Collaboration, Arzoumanian, Z., Brazier, A., et al.\ 2015, \apj, 813, 65 

\bibitem[Alam et al.(2021a)]{nano21a} Alam, M.~F., Arzoumanian, Z., Baker, P.~T., et al.\ 2021, \apjs, 252, 4. doi:10.3847/1538-4365/abc6a0

\bibitem[Alam et al.(2021b)]{nano21b} Alam, M.~F., Arzoumanian, Z., Baker, P.~T., et al.\ 2021, \apjs, 252, 5. doi:10.3847/1538-4365/abc6a1

\bibitem[{Nice \& Taylor(1995)}]{nt95}
Nice, D.~J. \& Taylor, J.~H. 1995, ApJ, 441, 429

\bibitem[Nice 
\& Sayer(1997)]{ns97} Nice, D.~J., \& Sayer, R.~W.\ 1997, \apj, 476, 261 

\bibitem[Nolan et al.(2012)]{naa+12} Nolan, P.~L., Abdo, 
A.~A., Ackermann, M., et al.\ 2012, \apjs, 199, 31 

\bibitem[Pellizzoni et al.(2009)]{ppp+09} Pellizzoni, A., 
Pilia, M., Possenti, A., et al.\ 2009, \apjl, 695, L115 

\bibitem[Pletsch et al.(2012)]{pga+12} Pletsch, H.~J., 
Guillemot, L., Allen, B., et al.\ 2012, \apj, 744, 105 

\bibitem[Pleunis et al.(2017)]{pleun17} Pleunis, Z., Bassa, C.~G., Hessels, J.~W.~T., et al.\ 2017, \apjl, 846, L19. doi:10.3847/2041-8213/aa83ff

\bibitem[Polzin et al.(2018)]{pbc+18} Polzin, E.~J., Breton, R.~P., Clarke, A.~O., et al.\ 2018, \mnras, 476, 1968. doi:10.1093/mnras/sty349


\bibitem[Radhakrishnan 
\& Cooke(1969)]{rc69} Radhakrishnan, V., \& Cooke, D.~J.\ 1969, \aplett, 3, 225

\bibitem[Ransom et al.(2002)]{rem02} Ransom, S.~M., 
Eikenberry, S.~S., \& Middleditch, J.\ 2002, \aj, 124, 1788

\bibitem[Ransom et al.(2011)]{rrc+11} Ransom, S.~M., Ray, 
P.~S., Camilo, F., et al.\ 2011, \apjl, 727, L16

\bibitem[Ray et al.(2011)]{ray11} Ray, P.~S., Kerr, M., Parent, D., et al.\ 2011, \apjs, 194, 17. doi:10.1088/0067-0049/194/2/17

\bibitem[Ray et al.(2012)]{rap+12} Ray, P.~S., Abdo, A.~A., 
Parent, D., et al.\ 2012, arXiv:1205.3089

\bibitem[Roberts et al.(2002a)]{rhr+02} Roberts, M.~S.~E., 
Hessels, J.~W.~T., Ransom, S.~M., et al.\ 2002, \apjl, 577, L19

\bibitem[Roberts et al.(2002b)]{rgr02} Roberts, M.~S.~E., 
Gaensler, B.~M., 
\& Romani, R.~W.\ 2002, Neutron Stars in Supernova Remnants, 271, 213

\bibitem[Roberts(2005)]{rob05} Roberts, M.~S.~E.\ 2005, 
Advances in Space Research, 35, 1142

\bibitem[Roberts(2011)]{rob11} Roberts, M.~S.~E.\ 2011, 
American Institute of Physics Conference Series, 1357, 127 

\bibitem[Roberts(2013)]{rob13} Roberts, M.~S.~E.\ 2013, IAU 
Symposium, 291, 127

\bibitem[Roberts et al.(2015)]{rmg+15} Roberts, M.~S.~E., 
McLaughlin, M.~A., Gentile, P.~A., et al.\ 2015, arXiv:1502.07208

\bibitem[Romani et al.(2021)]{rkf+21} Romani, R.~W., Kandel, D., Filippenko, A.~V., et al.\ 2021, \apjl, 908, L46. doi:10.3847/2041-8213/abe2b4

\bibitem[Ruderman et al.(1989)]{rst89} Ruderman, M., Shaham, 
J., \& Tavani, M.\ 1989, \apj, 336, 507

\bibitem[Russell et al.(2019)]{rbl+19} Russell, D.~M., Bramich, D.~M., Lewis, F., et al.\ 2019, Astronomische Nachrichten, 340, 278. doi:10.1002/asna.201913610

\bibitem[Sanpa-Arsa(2016)]{2016PhDT.......539S} Sanpa-Arsa, S.\ 2016, Ph.D. Thesis

\bibitem[Shannon 
\& Cordes(2010)]{sc10} Shannon, R.~M., \& Cordes, J.~M.\ 2010, \apj, 725, 1607 

\bibitem[Shklovskii(1970)]{s70} Shklovskii, I.~S.\ 1970, 
\sovast, 13, 562

\bibitem[Smedley et al.(2015)]{stf+15} Smedley, S.~L., Tout, C.~A., Ferrario, L., \& Wickramasinghe, D.~T.\ 2015, \mnras, 446, 2540 

\bibitem[Stappers et al.(2003)]{Stappers2003} Stappers, B.~W., 
Gaensler, B.~M., Kaspi, V.~M., et al.\ 2003, Science, 299, 1372 

\bibitem[Stappers et al.(2014)]{sah+14} Stappers, B.~W., 
Archibald, A.~M., Hessels, J.~W.~T., et al.\ 2014, \apj, 790, 39 

\bibitem[Stairs(2004)]{sta05} Stairs, I.~H.\ 2004, Science, 
304, 547

\bibitem[Stovall et al.(2014)]{slr+14} Stovall, K., Lynch, 
R.~S., Ransom, S.~M., et al.\ 2014, \apj, 791, 67

\bibitem[Strader et al.(2014)]{scs+14} Strader, J., Chomiuk, 
L., Sonbas, E., et al.\ 2014, \apjl, 788, LL27 

\bibitem[Strader et al.(2019)]{Strader19} Strader, J., Swihart, S., Chomiuk, L., et al.\ 2019, \apj, 872, 42. doi:10.3847/1538-4357/aafbaa

\bibitem[Swihart et al.(2022)]{Swihart22} Swihart, S.~J., Strader, J., Chomiuk, L., et al.\ 2022, \apj, 941, 199. doi:10.3847/1538-4357/aca2ac

\bibitem[Swihart et al.(2022)]{Swihart22b} Swihart, S.~J., Strader, J., Aydi, E., et al.\ 2022, \apj, 926, 201. doi:10.3847/1538-4357/ac4ae4

\bibitem[Tauris \& Savonije(1999)]{ts99} Tauris, T.~M., \& Savonije, G.~J.\ 1999, \aap, 350, 928

\bibitem[Tauris et al.(2011)]{Tauris2011} Tauris, T.~M., Langer, N., \& Kramer, M.\ 1999, \mnras, 416, 2130

\bibitem[Taylor(1992)]{t92} Taylor, J.~H.\ 1992, Royal 
Society of London Philosophical Transactions Series A, 341, 117

\bibitem[The Fermi-LAT Collaboration(2014)]{4CATel} The 
Fermi-LAT Collaboration 2014, ATel \#5838

\bibitem[Acero et al.(2015)]{3fgl} Acero, F., Ackermann, M., Ajello, M., et al.\ 2015, \apjs, 218, 23. doi:10.1088/0067-0049/218/2/23

\bibitem[Venter et al.(2009)]{vhg09} Venter, C., Harding, 
A.~K., \& Guillemot, L.\ 2009, \apj, 707, 800

\bibitem[van Haasteren et al.(2011)]{vanHaasteren2011} van Haasteren, R., Levin, Y., Janssen, G.~H., et al.\ 2009, \mnras, 414, 3117

\bibitem[Roy \& Bhattacharyya(2013)]{Roy13} Roy, J. \& Bhattacharyya, B.\ 2013, \apjl, 765, L45. doi:10.1088/2041-8205/765/2/L45

\bibitem[Wadiasingh et al.(2017)]{Wadiasingh17} Wadiasingh, Z., Harding, A.~K., Venter, C., et al.\ 2017, \apj, 839, 80. doi:10.3847/1538-4357/aa69bf

\bibitem[Wang et al.(2021)]{wlc+21} Wang, P., Li, D., Clark, C.~J., et al.\ 2021, Science China Physics, Mechanics, and Astronomy, 64, 129562. doi:10.1007/s11433-021-1757-5


\bibitem[Yao et al.(2017)]{ymw17} Yao, J.~M., Manchester, R.~N., \& Wang, N., 2017, \apj, 835, 29. doi:10.3847/1538-4357/835/1/29


\end{thebibliography}
\end{document}